\documentclass[journal]{new-aiaa}
\usepackage[utf8]{inputenc}
\usepackage{textcomp}

\usepackage{graphicx}
\usepackage[version=4]{mhchem}
\usepackage{siunitx}
\usepackage{longtable,tabularx}
\usepackage{hyperref}
\hypersetup{
    hidelinks
}

\usepackage{url}
\usepackage{subcaption}
\usepackage{enumitem}
\usepackage{amsmath}

\usepackage{amssymb}
\usepackage{amsfonts}
\usepackage{gensymb}

\usepackage{bm}

\usepackage{multirow}
\usepackage{ulem}

\usepackage[title]{appendix}
\usepackage{array}   
\usepackage{changes}
\usepackage{lipsum} 
\usepackage{xcolor} 
\usepackage{todonotes}

\usepackage{lineno} 

\usepackage[version=4]{mhchem}
\usepackage{siunitx}
\usepackage{longtable,tabularx}
\title{
Shape optimization for trailing-edge noise reduction using large-eddy simulation and ensemble-based method 
}

\author{Qingyong Luo \footnote{Ph.D. student, LNM, Institute of Mechanics; luoqingyong@imech.ac.cn }}

\author{Xin-Lei Zhang \footnote{Associate professor, LNM, Institute of Mechanics; zhangxinlei@imech.ac.cn (Co-corresponding author).}}

\author{Guowei He \footnote{Professor, LNM, Institute of Mechanics; hgw@lnm.imech.ac.cn (Corresponding author).}}

\affil{State Key Laboratory of Nonlinear Mechanics, Chinese Academy of Sciences, 100190 Beijing, \\ People’s Republic of China}

\affil{University of Chinese Academy of Sciences, 100049 Beijing, People’s Republic of China}

\begin{document}

\maketitle

\begin{abstract}
In this work, the trailing-edge shape of an airfoil is optimized to reduce the acoustic noise based on large-eddy simulation (LES).
It is achieved by the ensemble Kalman method, which can enhance the optimization efficiency by using the gradient of cost function approximated with sample covariances.
Moreover, the update scheme is reformulated to impose smoothness regularization and enable simultaneous reduction in the trailing edge noise and the drag-to-lift ratio.
The trailing edge is optimized with a reduced bevel angle based on the ensemble Kalman method.
The flow field near the optimal trailing edge shows that the flow separation and vortex shedding are suppressed compared to the baseline shape, indicating a significant decrease in the drag-to-lift ratio and noise generation.
Also, the spectral proper orthogonal decomposition method is used to analyze the flow structure around the trailing edge, identifying that the optimal shape achieves acoustic noise reduction by disrupting large-scale flow structures.
Further, the spectrum of Lighthill stress reveals that the optimal trailing edge suppresses the high-frequency noise through the nonlinear interaction of reduced low-frequency velocity fluctuations.
\end{abstract}

\section*{Nomenclature}

{
\renewcommand\arraystretch{1.0}
\noindent\begin{longtable*}{@{}l @{\quad=\quad} l@{}}
$\mathsf{a}$     &  shape parameters , i.e., dimensionless vertical displacement of control points   \\
$b$     & span of the trailing edge (m) \\

$C_d$   & drag coefficient \\
$C_f$   & friction coefficient \\
$C_l$   & lift coefficient \\
$C_p$   & pressure coefficient \\
$c_\infty$  & speed of sound at infinity (m/s) \\
$D$     & the drag-to-lift ratio \\
$\bm{F}$          & resultant force on the airfoil (N) \\
$f$ & temporal frequency (Hz) \\
$G$     & regularization term \\
$\mathcal{H}, \mathcal{D}$     &  model operator that maps the neural network weights to the observed quantities \\
$H$          &  acoustic power (Pa$^2$)  \\
$h$     & thickness of the airfoil (m)\\
$\bm{I}$ & unit tensor \\
$J$   & cost function \\
$\mathbf{K}$ & Kalman gain \\
$k$     & acoustic wave number (m$^{-1}$) \\
$L_s$  & length scale of turbulence fluctuation (m) \\
$L$    & chord length (m)\\
$\mathcal{M}$ & composite operator\\     
$M$             &  number of samples \\
$Ma$            &  Mach number    \\
$N$             &  maximum iteration number \\
$N_w$             &  number of weight \\
$\mathsf{P}$    &  model error covariance   \\
$p$          &  mean pressure (Pa)   \\
$p_a$ & pressure at far-field point (Pa)\\
$\mathsf{R}, \mathsf{Q}$    &  observation error covariance  \\
$R$     & radius of circular arc (m) \\
$Re$  & Reynolds number \\  
$r$ & radial distance (m)\\
$\mathbf{S}$ & mean strain rate (s$^{-1}$)\\
$S$ & acoustic source term (kg $\cdot$ m$^{1/2}$ $\cdot$ s$^{-2}$ )\\
$U$   & freestream velocity (m/s) \\
$u_r$   & radial velocity (m/s) \\
$u_s$  & velocity scale of turbulence fluctuation (m/s) \\
$u_\theta$  & tangential velocity (m/s) \\
$\mathbf{W}$     & weight of regularization term \\
$w$     & weight parameter of multi-objective cost functions\\
$\boldsymbol{x}$ & position vector of far-field point (m) \\
$\boldsymbol{y}$ & position vector of source-field point (m) \\
$y^+$     &  nondimensional wall distance  \\
SPL   & sound pressure level (dB) \\
TV    & total variation  \\
$\|*\|$       &  L2 norm  \\
$\hat{*}$ & Fourier coefficient \\
$\bar{*}$ & filter operation \\
$\langle * \rangle$ & ensemble mean \\
$\theta$ & polar angle \\
$\lambda$ & regularization parameter \\
${\mu, \mu _t }$  &  dynamic viscosity and turbulent viscosity (m$^2$/s)  \\
$\rho$ & density (kg/m$^3$) \\
$\Phi$  &  spectrum \\
$\omega$ & angular frequency (s$^{-1}$)  \\

\multicolumn{2}{@{}l}{Superscripts}\\
${j}$ & index of iteration step \\

$n$ & index of iteration \\
${\top}$ & transpose \\

$\prime$ & fluctuation \\

\multicolumn{2}{@{}l}{Subscripts}\\
$\text{0}$ & baseline \\
$\text{a}$ & augmented \\
$\text{r}$ & regularization \\
$\text{ref}$ & reference \\

${\infty}$    & at infinity \\

\end{longtable*}}

\section{Introduction}

Trailing edge noise represents a significant concern in aerodynamic design, as it constitutes the dominant source of high-frequency acoustic emissions from lift surfaces, such as aircraft wings and rotating blades~\citep{howe1999trailing,oberai2002computation}.
The generation of this noise stems from the turbulent boundary layer flows convecting over the trailing edge, scattering the surface pressure fluctuation due to the abrupt change of boundary condition.
During this process, the strong turbulent kinetic energy is converted into acoustic energy and radiated into the far field~\citep{lee2021turbulent}, which contributes significantly to overall noise emission.

Shape optimization offers a passive flow control approach to reduce trailing edge noise~\citep{choi2008control}.
It has been investigated that small geometric changes to a trailing edge can lead to substantial variations in the acoustic noise spectra~\citep{Blake2017,Blake2017Vol2}.
Geometric modifications such as serrants~\citep{howe1991aerodynamic}, brushes~\citep{finez2010broadband}, and finlet treatments~\citep{gstrein2023trailing} have been proposed to suppress the trailing edge noise, mainly by disrupting coherent flow structures near the trailing edge.
Given this, optimizing the trailing-edge shape is an effective way to improve the acoustic performance of airfoils.

Shape optimization of trailing edges requires scale-resolving simulations to evaluate the acoustic noise emission associated with geometric parameters.
The large-eddy simulation (LES) offers a widely validated and computationally efficient approach~\citep{wang2006computational,zhu_large-eddy_2022} for predicting trailing edge noise compared to the direct numerical simulation (DNS)~\citep{Sandberg2008}.
Specifically, the LES can resolve the most energetic flow scales relevant to turbulence noise generation, which are further used to estimate the trailing edge noise based on acoustic analogy~\citep{manoha2000trailing,oberai2002computation,ewert2004simulation,zhou_simplified_permeable_2021}.
In particular, \citet{wang2000computation} and \citet{manoha2000trailing} pioneered using incompressible large-eddy simulation and the acoustic analogy of the Ffowcs Williams and Hall to calculate the trailing edge noise.
Furthermore, \citet{wolf2012convective} combined the compressible large-eddy simulation and the Ffowcs Williams \& Hawkings acoustic analogy to predict the trailing edge noise from a NACA0012 airfoil.
These works make the far-field noise prediction in remarkable agreement with experimental measurements, demonstrating the predictive capability of the LES for trailing edge noise.

LES-based shape optimization faces difficulties with the widely used adjoint-based method~\citep{mohammadi2009applied} due to the chaotic effects of the turbulence.
Specifically, the adjoint-based method often provides divergent model gradients in chaotic systems, due to the severe sensitivity of time-averaged statistics to initial conditions~\citep{lea2000sensitivity}.
The least squares shadowing method~\citep{wang2014least} is proposed to alleviate the divergence issue by regularizing the ill-posed inverse problem with the closest trajectory to a pre-specified reference.
However, the adjoint-based method requires intensive effort in code development and maintenance of adjoint solvers.
Particularly when used for multi-objective optimization problems involving acoustic cost functions, specific adjoint solvers for each objective need to be developed.
Moreover, the adjoint-based method demands significant memory storage as it needs to restore the previously computed instantaneous flow fields at each time step for gradient calculation.
For this reason, gradient-free methods have emerged for LES-based shape optimization mainly due to their numerical robustness in chaotic problems.

The mesh adaptive direct search (MADS) method~\citep{audet2006mesh} is a gradient-free method that has been used for the LES-based shape optimization to reduce trailing edge noise.
This method is an extension of generalized pattern search that iteratively explores the space of design variables on a discretized mesh.
Moreover, it allows local exploration in a dense set of points with adaptive mesh refinement.
\citet{marsden2007trailing} demonstrated the effectiveness of the MADS method for reducing the trailing edge noise based on the large-eddy simulation.
This method has also been extended to different LES-based shape optimization problems, e.g., for increasing the lift-to-drag ratio of the SD7003 airfoil~\citep{karbasian2022gradient} and reducing the trailing edge noise of the NACA0012 airfoil~\citep{hamedi2023gradient}.
However, the method remains challenging for high-dimensional optimization problems, as the search space grows exponentially with the number of geometric parameters to be optimized.

The ensemble Kalman method~\citep{evensen2009data,iglesias2013ensemble} is a stochastic optimization method that can provide gradient approximation with random samples without requiring the adjoint-based gradient.
It is achieved by estimating the gradient of the cost function with the sample covariance of geometric parameters and LES predictions, which is further used to guide the optimization search direction.
With the ensemble-based gradient approximation, one can provide relatively high optimization efficiency compared to the direct search method.
This method has been employed for high-dimensional inverse problems~\citep{kovachki2019ensemble}, such as learning neural network-based turbulence closures from velocity data~\citep{zhang_ensemble-based_2022,liu2023learning,villiers2025enhancing}, inferring optimal Reynolds stress fields from sparse observation~\citep{xiao_quantifying_2016,zhang2022assessment}, and optimizing modeled quantities of jet flows based on far-field noise data~\citep{zhang2022acoustic}.
The variant of the ensemble Kalman method has also been introduced to shape optimization in laminar flows with upstream noises~\citep{lorente2023shape} and transitional boundary layer flows~\citep{jahanbakhshi2023optimal}.
Furthermore, the regularized ensemble Kalman method~\citep{zhang2020regularized} was introduced for the LES-based shape optimization problem~\citep{zhang2024large} in turbulent flows, which allows imposing the smoothness regularization and improving the optimization efficiency by utilizing the ensemble-based gradient and Hessian.
However, the method has only been demonstrated to mitigate the unsteadiness of turbulent wakes in flows with low Reynolds numbers.
It is worthy of further investigation for acoustic shape optimization in flows with high Reynolds numbers.

The acoustic shape optimization based on the LES faces additional difficulties.
On the one hand, the trailing-edge shape that reduces acoustic noise may deteriorate the aerodynamic performance, e.g., increasing drag and reducing lift forces.
This necessitates multi-objective optimization involving the drag-to-lift ratio to enhance both the acoustic and aerodynamic performance of the optimized trailing edge.
On the other hand, the LES-based acoustic prediction needs significant computational cost~\citep{bres2019modelling} as the grid refinements are required to capture the turbulent fluctuations near the trailing edge that lead to high-frequency noise generation.
Hence, the ensemble Kalman method needs further development to tackle multi-objective optimization problems with improved computational efficiency.

In this work, we present the regularized ensemble Kalman method~\citep{zhang2020regularized} to optimize trailing-edge shape based on large eddy simulation.
In order to address the issues of the LES-based acoustic shape optimization, the method is reformulated to minimize a multi-objective function concerning both the drag-to-lift ratio and the acoustic noise power based on the ensemble-based gradient approximation.
Also, the regularization step is modified for robust enforcement of the smoothness constraint.
Moreover, the weighted sum method is employed to find the Pareto frontier and determine the optimal weight parameters to balance the acoustic and aerodynamic objectives.
In addition, the zonal LES~\citep{wang2000computation,TUCKER2004267} is used to facilitate computational efficiency, which integrates the RANS method for the evaluation of the drag-to-lift ratio and the LES method for the prediction of acoustic noise.
As such, the ensemble Kalman method and the zonal LES are integrated to optimize the trailing-edge shape efficiently, demonstrating its capability for acoustic noise reduction at flows with high Reynolds numbers.

The rest of this paper is outlined as follows.
The shape optimization problem for trailing edge noise reduction is formulated in Section~\ref{sec:problem}.
The regularized ensemble Kalman method for multi-objective shape optimization is illustrated in Section~\ref{sec:method}.
The optimization process with the ensemble-based method is presented in Section~\ref{sec:results}.
The flow field around the optimal trailing-edge shape is discussed in Section~\ref{sec:Optimal}.
Finally, the paper is concluded in Section~\ref{sec:conclusion}.

\section{Problem formulation}
\label{sec:problem}

\subsection{Trailing edge of an airfoil}

\begin{figure}[!htb]
    \centering
    \includegraphics[width=1.0\linewidth]{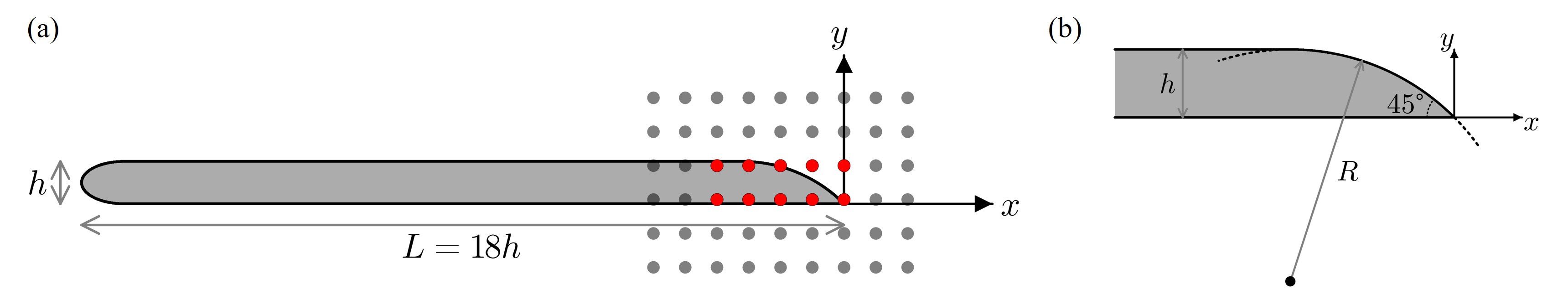}
    \caption{(a) Geometry of airfoil with the $45\degree$ beveled trailing edge. The thickness of the airfoil is $h$, and the chord length $L$ is $18h$. (b) Zoomed view of trailing edge.} \label{fig:geometry}
\end{figure}

An airfoil with a beveled~$45\degree$ trailing edge and a chord-to-thickness ratio of $18$ is investigated for acoustic shape optimization in this work.
The flow configuration is presented in Figure~\ref{fig:geometry}, where the thickness of the airfoil is $h$.
The upper surface of the trailing-edge shape is a smoothly transitioned circular arc, as shown in Figure~\ref{fig:geometry}(b).
The radius of the circular arc can be calculated through the geometric relationship as $R=h/(1-\cos{(\pi/4)})$.
The Reynolds number based on chord length $L$ and freestream velocity $U$ is $ Re = 1.9 \times 10^6$, and the freestream Mach number is $Ma=0.09$.
This airfoil geometry has been extensively investigated by the experimental measurements~\citep{blake1975statistical,olson2004experimental} and large-eddy simulation~\citep{wang2005computation}, which can be used to validate the predictive accuracy of the present LES.
Also, the gradient-free optimization approach, i.e., mesh adaptive direct search (MADS) method, has been applied to optimize the trailing-edge geometry~\citep{marsden2007trailing}.
The optimal shape with the MADS method can be used to validate the effectiveness of the proposed ensemble-based optimization method, which is presented in Appendix~\ref{sec:validate}.

Shape deformation is needed to update the geometry of the airfoil based on optimized geometric parameters.
Here it is achieved with the volumetric B-spline interpolation method~\citep{MARTIN201426}, given the prescribed position of control points.
As depicted in Figure~\ref{fig:geometry}(a), a deformable region is prescribed around the trailing edge, ranging from $-4.5h$ to $1.5h$ in the streamwise direction and from $-1.5h$ to $2.5h$ in the vertical direction.
Within this region, the positions of the grid points can be varied with a set of control points.
Among these control points, the 10 red control points are designated as movable, and their positions can be adjusted during the optimization process.
The gray control points are designated as immovable, and their positions remain fixed to ensure grid continuity between the deformable and non-deformable regions of the mesh.
The volumetric B-spline basis functions are determined recursively based on polynomial orders~\citep{piegl2012nurbs}.
Further, the position of each grid point within the deformable region is interpolated based on the updated positions of control points.
In this work, the dimensionless vertical displacement of the red movable control points, normalized by $h$, is used as the geometric parameters~$\mathsf{a}$ to be optimized.

\subsection{Cost function formulation} \label{subsec:cost_fun}

Multi-objective shape optimization is often required for trailing edge noise reduction, as the optimal shape for noise reduction can deteriorate the aerodynamic performance, e.g., in the coefficients of lift~$C_l$ and drag~$C_d$.
Given this, we aim to minimize both the acoustic noise and the drag-to-lift ratio of the airfoil.
To achieve this goal, the cost function is formulated as
\begin{equation}
   \min  J_0 = \| \mathcal{H}[\mathsf{a}] \|_\mathbf{R}^2 + \| \mathcal{D}[\mathsf{a}] \|_\mathbf{Q}^2 \text{,}
\end{equation}
which involves two objectives, i.e., the total acoustic power at the far field, represented by $\mathcal{H}[\mathsf{a}]$, and the drag-to-lift ratio ($C_d/C_l$), represented by $\mathcal{D}[\mathsf{a}]$. Both terms are defined as weighted squared norms, with $\mathbf{R}$ and $\mathbf{Q}$ serving as the corresponding weighting matrices.
For the conciseness, the expression $\| \boldsymbol{v} \|^2_\mathbf{A}$ is used to indicate a weighted squared norm, calculated as $\boldsymbol{v}^\top \mathbf{A}^{-1} \boldsymbol{v}$ for a vector $\boldsymbol{v}$ with weight matrix $\mathbf{A}$. 
For the one-dimensional quantity as considered here, the objective is a scalar, and the weighted norm $\| \boldsymbol{v} \|^2_\mathbf{A}$ is calculated as $v^2/A$.

The drag-to-lift ratio $\mathcal{D}[\mathsf{a}]$ concerning the shape parameter can be computed based on the mean pressure~$p$ and shear stress along the airfoil surface as
\begin{equation}
    \begin{aligned}
        & \mathcal{D}[\mathsf{a}] = \frac{C_d }{C_l} = \frac{|F_x|}{|F_y|} \text{,} \\
        & \mathbf{F} = \oint (-p \mathbf{I} + 2 \mu \mathbf{S})\cdot d\mathbf{A} \text{.}
    \end{aligned}
    \label{eq:Da}
\end{equation}
In the above formula, $\mathbf{F}$ is the resultant force exerted by the fluid on the airfoil, with $F_x$ and $F_y$ representing its horizontal and vertical components, respectively.
$\mathbf{I}$ is the unit tensor, $\mu$ is the molecular viscosity, and $\mathbf{S}$ is the mean strain rate.
The aerodynamic force exerted on the airfoil can be obtained by solving the Reynolds-averaged Navier-Stokes (RANS) equations.

\begin{figure}[!htb]
    \centering
    \includegraphics[width=1.0\linewidth]{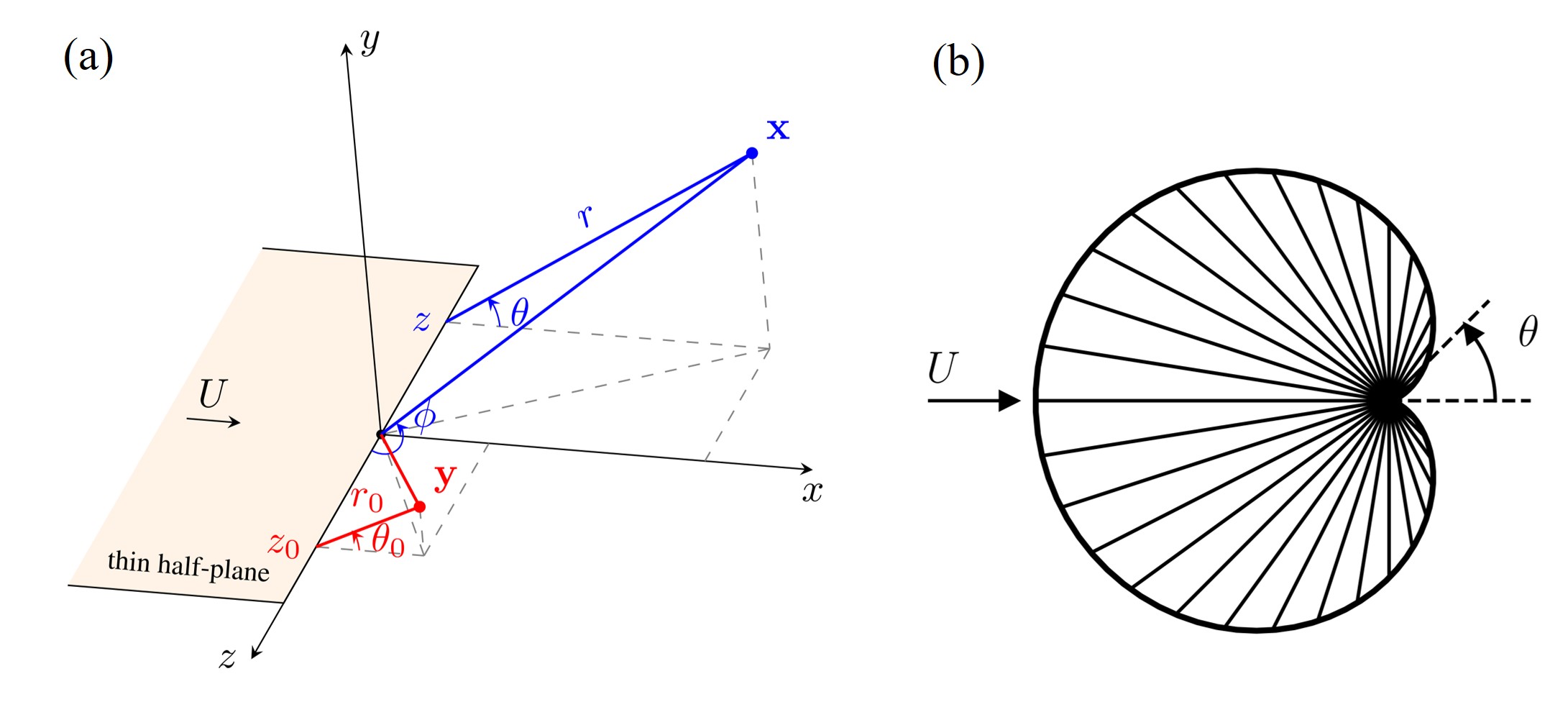}
    \caption{(a) Definition of the cylinder-polar coordinate system for the half-plane Green's function solution. (b) The cardioid directivity of trailing edge noise.} \label{fig:coordinates}
\end{figure}

The trailing edge noise is estimated with the LES-predicted instantaneous velocity fields in conjunction with the acoustic analogy of Ffowcs Williams and Hall in this work.
Specifically, the flow field is first obtained using the LES method to provide the predictions of the Lighthill stress.
Further, the pressure fluctuation at the far field point~$\mathbf{x}$ can be obtained in the frequency domain based on the LES-predicted Lighthill stress and the thin half-plane Green's function as~\citep{williams1970aerodynamic}
\begin{equation}
    \hat{p}_a(\mathbf{x},\omega) \approx \frac{e^{i(k|\mathbf{x}|-\pi/4)} }{2^\frac{5}{2} \pi^\frac{3}{2} |\mathbf{x}| } ( k \sin{\phi} )^\frac{1}{2} \sin{\frac{\theta}{2}} \hat{S}(\omega) \text{,}
    \label{eq:pa}
\end{equation}
where $\omega$ is the angular frequency, $k$ is the acoustic wave number, and $\hat{S}(\omega)$ is the Fourier coefficient of $S(t)$ in the noise source region~$\Omega$ with
\begin{equation}
    S(t) = \int_\Omega \frac{\rho}{r_0^{3/2}}\left[ (u_\theta^2 - u_r^2) \sin{\frac{\theta_0}{2}} - 2u_r u_\theta \cos{\frac{\theta_0}{2}} \right] d^3\mathbf{y} \text{.}
    \label{eq:St}
\end{equation}
The position vectors $\mathbf{x}(r,\theta,z)$ and $\mathbf{y}(r_0,\theta_0,z_0)$ represent the far-field and source-field points, respectively, which are defined in the cylinder-polar coordinate system as shown in Figure~\ref{fig:coordinates},  with $\sin{\phi} = r/|\mathbf{x}|$.
The velocity $u_r$ and $u_\theta$ are the radial and tangential components of the velocity vector in cylindrical coordinates, respectively.
Further, the power spectrum of acoustic pressure can be derived as
\begin{equation}
    \Phi_{p_a}=|\hat{p}_a|^2 = \frac{\omega \sin ^2 \frac{\theta}{2} \sin \phi}{2^5 \pi^3|\mathbf{x}|^2 c_{\infty}}\Phi_{s s}(\omega)  \text{,}
    \label{eq:Phi_pa}
\end{equation}
where $\Phi_{ss}=|\hat{S}(\omega)|^2$ is the spectrum of the acoustic source term, and $c_\infty$ is the speed of sound in the incoming flow at infinity. 
The cardioid directivity pattern of the trailing edge noise is depicted in Figure~\ref{fig:coordinates}(b), with $\sin^2{\left(\theta/2 \right)}$ plotted in the polar coordinate.
It can be observed that the maximum acoustic energy is propagated along the upstream direction, while there is no acoustic energy in the downstream direction.
By following the work of~\citet{marsden2007trailing}, we define the objective $j(\omega)$ of far-field noise as
\begin{equation}
    j(\omega) \equiv \Phi_{p_a}\left( \frac{2^5 \pi^3 |\mathbf{x}|^2c_{\infty}}{\sin ^2 \frac{\theta}{2} \sin \phi}\right) = \omega \Phi_{s s}(\omega)  \text{,}
    \label{eq:jw}
\end{equation}
which is proportional to the far-field pressure spectrum~$\Phi_{p_a}$ at a given frequency~$\omega$.
The acoustic cost function~$\mathcal{H}[\mathsf{a}]$ is defined by integrating the acoustic spectrum over entire frequencies as
\begin{equation}
    \mathcal{H}[\mathsf{a}] = \int j(\omega) d\omega \text{.}
    \label{eq:Ha}
\end{equation}
Based on Eq.~\eqref{eq:jw}, the cost functional value is proportional to the total acoustic power at the far-field observation location $\mathbf{x}$.

Note that the acoustic analogy of Ffowcs Williams and Hall for trailing-edge noise prediction relies on several assumptions. 
First, viscous and nonisentropic contributions in Lighthill stress are neglected due to the high Reynolds number and low Mach number.
Also, Lighthill's equation is solved using the thin half-plane Green's function, which is effective for describing the scattered field in the acoustically thin foil limit ($kh \ll 1$), or when the thickness $h$ is much smaller than the acoustic wavelength~\citep{crighton1971scattering}.
Moreover, given the slender airfoil assumption, the directivity of sound propagation is consistent with that shown in Figure~\ref{fig:coordinates}(b), independent of the optimized shape variations.
In addition, Eq.~\eqref{eq:pa} is derived given a source region smaller than one acoustic wavelength with a compact spanwise length ($kr_0 \ll 1$).
However, the noise source region $\Omega$ should be sufficiently large to ensure the convergence of the integral solution. 
In light of this issue, the influence of the region size on the noise prediction is investigated in this work as presented in Appendix~\ref{sec:source region}.
Based on the results, we set the noise source region with a radius of $R_0=2.6h$ given the convergence and accuracy of the integral solution. 
Extensive discussions and validations~\citep{wang2000computation,wang2005computation,wang2006computational} have demonstrated the feasibility of this approach in calculating trailing-edge noise, provided that the airfoil is long and thin relative to the acoustic wavelengths~($ h \ll 2\pi/k \ll L $) as investigated in this work.

\subsection{Cost function evaluation with zonal large eddy simulations}

In order to improve the computational efficiency, the zonal large-eddy simulation~\citep{wang2000computation} is used to evaluate the aerodynamic and acoustic performance of the airfoil.
Specifically, the RANS simulation is first conducted over the entire computational domain to provide the predictions of mean aerodynamic force.
The RANS-predicted Reynolds stress at the zone interface is used to generate inlet turbulence for the LES through the divergence-free synthetic eddy method(DFSEM)~\citep{poletto_new_2013}.
Further, the LES is conducted around the trailing edge to predict the acoustic noise by combining with the acoustic analogy of Ffowcs Williams and Hall.
Such hybrid strategies have been used for trailing edge noise prediction\cite{wang2000computation,marsden2007trailing}.
This approach utilizes a RANS simulation for the entire computational domain to determine the total aerodynamic forces (lift and drag).
Meanwhile, the LES is applied specifically around the trailing edge to predict the far-field noise spectra.

\begin{figure}[!htb]
\includegraphics[width=1.0\linewidth]{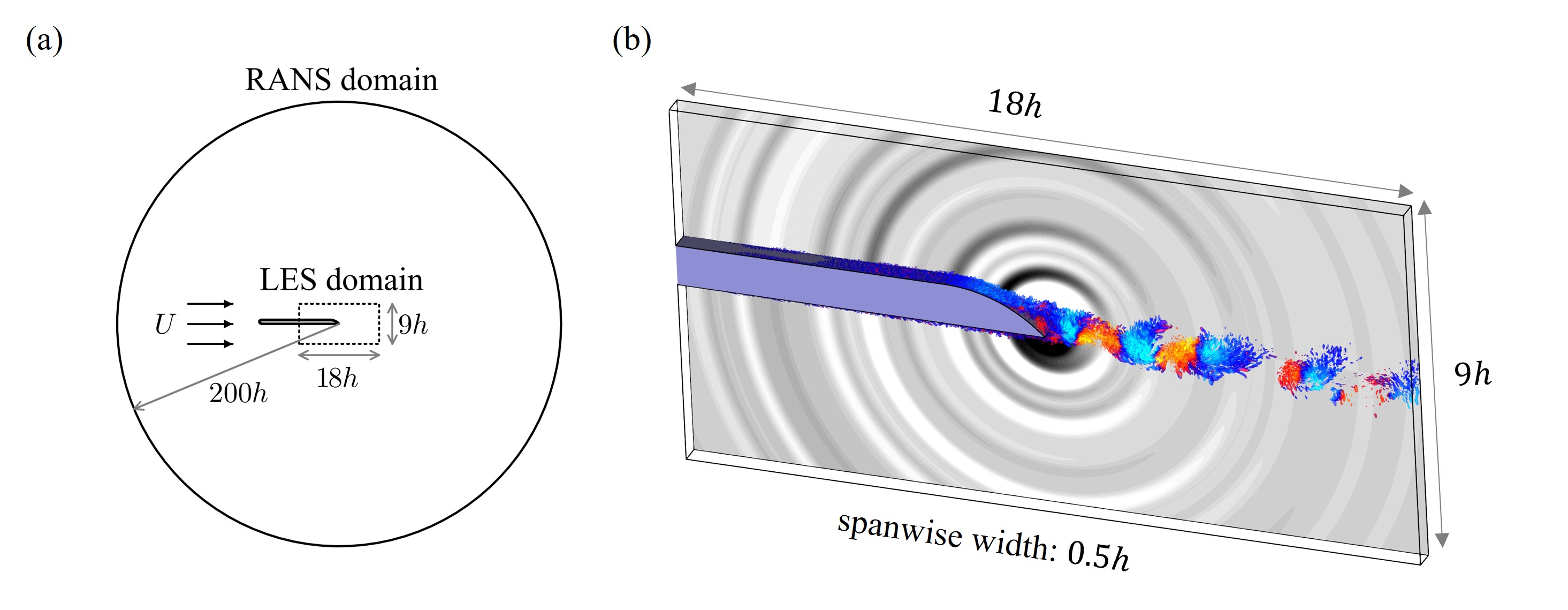}
        \caption{
        (a) Computational domain.
        (b) Q-criterion and radiated pressure fluctuations using Eq.~\eqref{eq:pa} from the LES results.} \label{fig:computation_domains}
\end{figure}

The computational domain of the zonal large-eddy simulation is presented in Figure~\ref{fig:computation_domains}(a).
The RANS simulation is performed over the entire domain, and the incompressible LES is conducted around the trailing edge with $18h$ in the streamwise direction, $9h$ in the vertical direction, and $0.5h$ in the spanwise direction.
The maximum radius of the computational domain is set to $200h$, which is sufficiently large to prevent inappropriate interference from the boundary conditions on the near-field flow.
Figure~\ref{fig:computation_domains}(b) presents the flow field from the LES results.
The iso-surface of the Q-criterion~\citep{hunt1988eddies} is colored by normal velocity, with the background showing the radiated pressure fluctuation.
It is noted that the employed incompressible LES inherently does not capture acoustic waves and their propagation. 
In Figure~\ref{fig:computation_domains}(b), a schematic visualization of far-field pressure fluctuations synthesized using Eq.~\eqref{eq:pa} is presented.

The incompressible RANS equations are solved to provide the prediction of the aerodynamic lift and drag.
The widely used Spalart-Allmaras (SA) turbulence model~\citep{spalart1992one-equation} is used for the Reynolds stress closure in this work.
The structured C-type grids are employed for the near-wall regions of the RANS simulations, while the unstructured grids are utilized for the outer regions.
The total number of grid cells is approximately $63,000$ cells.
The first off-wall height is $\Delta y^+ \approx 2$.
The solid wall is imposed with the no-slip boundary condition, and the outer boundary of the computational domain is treated with a freestream condition.
The RANS-predicted lift and drag coefficients are validated by comparing with the zonal LES predictions for the trailing edge section ($x/h>-9$).
The RANS-predicted lift and drag coefficients are $0.285$ and $0.0272$, respectively, which are very close to the zonal LES results, i.e., $0.269$ and $0.0283$.
It is noted that the SA model is a one-equation eddy-viscosity model, which does not explicitly provide the Reynolds stress. 
In this work, the Reynolds stresses are required to generate turbulent inlet condition for LES using the DFSEM.
It is computed using the mean strain rate tensor ($\mathbf{S}$) and the turbulent eddy viscosity ($\nu_t$) as
$\mathbf{\tau} = 2\nu_t \mathbf{S} - \frac{2}{3} k \mathbf{I}$, 
where $\mathbf{I}$ is the identity tensor, and $\nu_t$ is the turbulent eddy viscosity by the SA model.
The SA model does not provide the turbulent kinetic energy ($k$), and here it is estimated using the Bradshaw assumption, which is commonly used for the TKE estimate with the SA model~\cite{rung2003restatement,bourgoin:tel-02310681}.

As for the large-eddy simulation, the filtered incompressible Navier-Stokes equation is solved to predict the instantaneous flow fields, coupling with the dynamic Smagorinsky subgrid-scale stress model~\citep{germano1991dynamic,lilly1992proposed}. 
The mesh grid for large eddy simulation is $1500 \times 96 \times 48$, with around 7 million cells.
The streamwise and spanwise grid spaces measured in wall units are $\Delta x^+ \approx 48$ and $\Delta z^+ \approx 42$, respectively.
The first off-wall height is $\Delta y^+ \approx 2$.
Such computational domain size and grid resolution are adequate to accurately predict the broadband noise of the investigated airfoil based on the work of~\citet{wang2000computation}.
We also test a refined mesh with $\Delta y^+ \approx 1$.
The predicted SPL results are similar to the case of $\Delta y^+ \approx 2$, and the plots are omitted for brevity.
This further confirms that the resolution of $\Delta y^+\approx2$ is sufficient for acoustic prediction in this case.
The zonal LES strategy is employed to efficiently simulate the flow, which leverages the RANS solution to provide appropriate boundary conditions for the LES domain near the trailing edge. 
Specifically, the top and bottom boundaries of the LES domain are prescribed with the mean velocity components obtained from the steady RANS solution.
The inflow boundary of the zonal LES domain is imposed with synthetic turbulent fluctuations, which are generated using the DFSEM based on the mean velocity profile and the modeled Reynolds stress from the RANS solution. 
In the spanwise direction, periodic boundary conditions are applied. 
A convective outflow condition is applied at the outlet boundary of the LES domain.

\begin{figure}[!htb]
    \centering
    \includegraphics[width=0.95\linewidth]{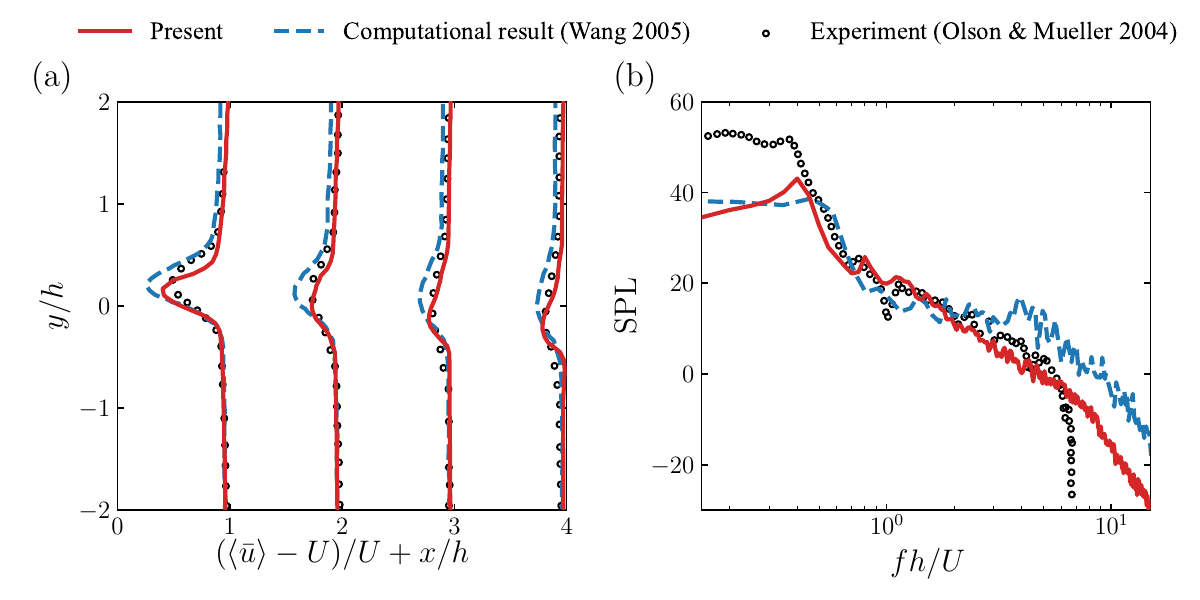}
    \caption{
    Validation of the present LES in the (a) mean velocity and (b) sound pressure spectrum by comparison with both computational results~\citep{wang2005computation} and experimental data~\citep{olson2004experimental}.
    }
    \label{fig:validate}
\end{figure}

The validation of the LES is performed through a comprehensive comparison with the LES results of~\citet{wang2005computation} and the experimental measurement of~\citet{olson2004experimental}. 
The comparison results are shown in Figure~\ref{fig:validate}. 
Figure~\ref{fig:validate}(a) compares velocity profiles in the wake between the present LES and the reference data.
The profiles present the variation of normalized streamwise velocity $\langle \bar{u}\rangle/U$ across the normalized vertical coordinate~$y/h$.
The velocity profiles show the wake development downstream of an airfoil, exhibiting a central region of reduced velocity.
The low-velocity region becomes wider downstream, indicating the wake spreading due to diffusion effects.
The present LES results agree well with the experimental data in terms of both the peak velocity decay at the center and the wake evolution along the streamwise direction.
The sound pressure levels (SPL) are computed to assess the capability of the present LES in noise prediction based on $\text{SPL} = 10 \log{\left( \frac{|\hat{p}_a|^2}{p_{\text{ref}}^2}\right)}$ with the reference pressure $p_{\text{ref}} = 20 \mu \text{Pa}$.
Figure~\ref{fig:validate}(b) presents a comparison of the predicted SPL with the reference data at the far-field position of $x/h=3$ and $y/h=21$.
The peaks of the SPL spectrum occur at low frequencies around~$fh/U=0.4$ and then decrease as frequency increases.
The predicted SPL shows good agreement with both computational and experimental references across a broad range of frequencies.
It can be observed that there exhibit notable differences from the measurement data in the high-frequency range, which is because the high-frequency noise is cut off by the low-pass filter in the experiment~\citep{olson2004experimental}.
As for the difference in low-frequency noise calculations, it is primarily attributed to significant near-field effects in the experiment. 
That is, when the acoustic wavelength is larger than the distance between the observation position and trailing edge,~i.e.,~$fh/U<h/(M|\mathbf{x}|) \approx 0.52$, the experimental results are contaminated by near-field pressure fluctuations~\citep{wang2005computation}.
Moreover, the approximation with the half-plane Green's function is invalid for low-frequency acoustic waves, which can result in prediction discrepancies as observed in Figure~\ref{fig:validate} according to~\citet{howe2001edge}.
In general, both the velocity profiles and sound pressure spectrum exhibit good agreement with the experimental data, demonstrating the predictive accuracy of the present LES in capturing the flow and noise characteristics of the airfoil.

\section{Multi-objectives optimization} 
\label{sec:method}

In multi-objective optimization problems, different objectives may contradict each other, leading to the optimal shape that reaches the least noise emission while increasing the aerodynamic drag.
In view of this, the weighted cost function should be formulated and minimized to find the optimal solution with the best compromise of the two objectives.
Genetic algorithms~\cite{holland1992genetic} are a widely recognized and commonly used method for multi-objective optimization~\cite{waschkowski2022multi}. 
However, as a heuristic algorithm, the method is often computationally prohibitive when objective function evaluations are costly.
For this reason, to the authors' knowledge, few works use genetic algorithms for LES-based shape optimization, due to substantial computational resources for each design evaluation.
The ensemble-based method offers a relatively efficient alternative for multi-objective optimization.
Here the method is employed to minimize the weighted cost function associated with the acoustic noise and the drag-to-lift ratio.
In the following, the weighted cost function and the ensemble Kalman method are illustrated in detail.

\subsection{Weighted cost function for Pareto frontier}

In multi-objective optimization, the Pareto frontier represents the set of optimal solutions where improving one objective inevitably degrades at least one other.
The weighted sum method~\citep{de2004multiobjective,marler2010weighted} is one widely used approach to approximate the Pareto frontier by integrating multiple objectives into one single cost function.
This method leverages the computational efficiency of single-objective optimization while allowing for the exploration of trade-offs between conflicting objectives.

The weighted cost function is formulated as a linear combination of multiple objective functions, each scaled by a corresponding weight.
Given $ n $ objective functions $J_1[\mathsf{a}], J_2[\mathsf{a}], \ldots, J_n[\mathsf{a}]$, the weighted cost function $J[\mathsf{a}]$ can be expressed as
\begin{equation}    
    J_w[\mathsf{a}] = \sum_{i=1}^{n} w_i J_i[\mathsf{a}] \text{,}
    \label{eq:wighted_sum_obj}
\end{equation}
where $w_i$ represents the weight assigned to the $i$-th objective function, ensuring that all weights are positive and sum up to 1.
The minimum of Eq.~\eqref{eq:wighted_sum_obj} is Pareto-optimal~\citep{zadeh1963optimality}. 
The optimal $\mathsf{a}^*$ depends on the weights $\mathbf{w} = [w_1, w_2,...,w_n]$, as
\begin{equation}
    \begin{aligned}
        \mathsf{a}^*(\mathbf{w}) = \arg \min_{ \mathsf{a} } J_w[\mathsf{a}] \quad s.t. \quad w_i >0 \text{,} \sum_{i=1}^{n}w_i=1 \text{.}
    \end{aligned}
\end{equation}
By scanning the weight parameters typically sampling uniformly, the obtained set of optimal solutions is approximately the Pareto frontier.
The weighted sum method with parameter scanning can provide all Pareto optimal points if the multi-objective problem is convex~\citep{marler2010weighted}.
While the method may struggle to fully capture the complete Pareto-optimal set for non-convex problems, it nonetheless offers a straightforward and computationally efficient means to approximate segments of the Pareto frontier.
Moreover, this method is capable of identifying an optimal set of weights that yields substantial enhancements across multiple objectives simultaneously.

In this work, the weighted cost function associated with the acoustic power~$\mathcal{H}[\mathsf{a}]$ and the drag-to-lift ratio~$\mathcal{D}[\mathsf{a}]$ is formulated as 
\begin{equation}
    \begin{aligned}
        J_w[\mathsf{a}] &= w J_1[\mathsf{a}] + (1-w) J_2[\mathsf{a}] \text{, with} \\
        J_1[\mathsf{a}] &= \| \mathcal{H}[\mathsf{a}] \|^2_{\mathbf{R}} \quad \text{and} \quad
        J_2[\mathsf{a}] = \| \mathcal{D}[\mathsf{a}]  \|^2_{\mathbf{Q}} \text{,}
    \end{aligned}
\end{equation}
where the weight parameter $w$ needs to be scanned to find the Pareto-optimal points.
The weighting parameters $\mathbf{R}$ and $\mathbf{Q}$ allow for normalizing the magnitudes of the two objectives to a comparable scale. 
The scalar weight $w \in [0, 1]$ (and $1-w$) is introduced to have the trade-off between the acoustic performance objective ($J_1$) and the aerodynamic efficiency objective ($J_2$) within the multi-objective optimization framework.
We adopt the uniform weight allocation method to find the Pareto frontier by uniformly sampling the weight parameters in this work. 
Specifically, we select weights by sampling from $0$ to $1$ with a fixed interval for the two objectives.
After obtaining the Pareto-optimal points, we select the optimal solution from the Pareto frontier with the least squared Euclidean distance of the objective vector to the origin, as
\begin{align}
    w^* = \arg \min_{w} \left[ \left( J_1[\mathsf{a}^*(w)] \right)^2 + \left( J_2[\mathsf{a}^*(w)] \right)^2 \right] \text{.}
    \label{eq:criterion}
\end{align}
With this criterion, we aim to determine an optimal shape that achieves the best compromise between the two objective quantities.
The optimal shape with the weight parameter~$w^*$ is used for the detailed flow analysis in this work.

\subsection{Ensemble Kalman method for multi-objective optimization}

The ensemble Kalman method~\citep{iglesias2013ensemble,evensen2018analysis} is employed to reduce the far-field acoustic power and the drag-to-lift ratio simultaneously by optimizing the trailing-edge shape.
The cost function includes both the far-field acoustic power and the drag-to-lift ratio of the airfoil as depicted in Subsection~\ref{subsec:cost_fun}.
Moreover, a regularization term~$\mathcal{G}[\mathsf{a}]$ is introduced to constrain the optimization process and ensure smoothness of the optimized shape~\citep{zhang2020regularized}. 
The weighted cost function with the regularization terms is 
\begin{equation}
    J = w \| \mathcal{H}[\mathsf{a}^{j+1}] \|^2_{\mathbf{R}} + (1-w)\| \mathcal{D}[\mathsf{a}^{j+1}]  \|^2_{\mathbf{Q}} + \lambda \| \mathcal{G}[\mathsf{a}^{j+1}]\|^2_{\mathbf{W}} + \|\mathsf{a}^{j+1}-\mathsf{a}^{j}\|^2_{\mathbf{P}} \text{.} 
    \label{eq:loss}
\end{equation}
The first two terms represent the weighted contributions from the far-field acoustic power and the drag-to-lift ratio of the airfoil, respectively.
The third term $\| \mathcal{G}[\mathsf{a}^{j+1}]\|^2_{\mathbf{W}}$ is the regularization term to ensure the smoothness of the optimized shape, with $\mathbf{W}$ being the weighting matrices.
The last term is introduced to prevent large updates between successive iterations, thereby promoting stability and convergence of the optimization process.
The weight matrix~$\mathbf{P}$ is obtained with the sample covariances of $\mathsf{a}$.
In this work, $\mathcal{G}[\mathsf{a}]$ is chosen as the difference of total variation between the updated trailing-edge profile~$l[\mathsf{a}]$ and the baseline profile~$l[\mathsf{a}_0]$, i.e.,
\begin{equation}
    \begin{aligned}
        & \mathcal{G}[\mathsf{a}] = \max(\text{TV}[\mathsf{a}] - \text{TV}[\mathsf{a}_0], \; 0) \text{,} \\
        & \text{TV}[\mathsf{a}] \equiv \frac{1}{h} \int_{l[\mathsf{a}]} \left| \frac{dy^l}{dx^l}\right| dx \text{,}
    \end{aligned}
\end{equation}
where $dy^l/dx^l$ indicates the slope of the profile~$l[\mathsf{a}]$ at the point $(x^l,y^l)$.
This smoothness regularization can avoid the total variation of the optimal shape larger than the baseline shape.
Note that the baseline geometric parameter $\mathsf{a}_0$ is equal to $\mathsf{0}$, indicating that all control points have no displacement.
For a discrete representation of the trailing edge profile $l[\mathsf{a}]$, the total variation can be approximated as
\begin{equation}
    \text{TV}[\mathsf{a}] \approx \sum_{i=1}^{n_l-1} \frac{\left| y_{i+1}^l - y_i^l \right|}{h} \text{,}
    \label{eq:d_TV}
\end{equation}
where $y_i^l$ is the vertical coordinate of the points of the trailing edge surface with $i=1,2,...,n_l$, and $n_l$ denotes the total number of discrete points.
This discrete form sums the absolute differences in the $y$-coordinates of adjacent points along the trailing edge profile, which can effectively approximate the total variation and measure the smoothness of the geometric shape.

Given the weight parameter $w$, the cost function involving the two objective terms can be expressed in the augmented form as
\begin{equation}
\begin{aligned}
    & J = \| \mathcal{M}[\mathsf{a}^{j+1}] \|^2_{\mathbf{R_a}} + \lambda \| \mathcal{G}[\mathsf{a}^{j+1}]\|^2_{\mathbf{W}} + \|\mathsf{a}^{j+1}-\mathsf{a}^{j}\|^2_{\mathbf{P}} \text{, with} \\
    & \mathcal{M}[\mathsf{a}] = \begin{bmatrix}
    \sqrt{w}\mathcal{H}[\mathsf{a}] \\
     \sqrt{1-w} \mathcal{D}[\mathsf{a}]
    \end{bmatrix}
    \quad \text{and} \quad 
    \mathbf{R_a} = 
    \begin{bmatrix}
    \mathbf{R} & 0\\
    0 & \mathbf{Q} 
    \end{bmatrix} \text{.}
\end{aligned} \label{eq:loss_aug}
\end{equation}
The shape parameters can be updated by minimizing the Eq.~\eqref{eq:loss_aug} as~\citep{zhang2020regularized}
\begin{equation}
    \begin{aligned}
        &\tilde{\mathsf{a}}^{j} =  \mathsf{a}^{j} - \lambda\mathbf{P}(\mathcal{G}^\prime[\mathsf{a}^{j}])^\top \mathbf{W}^{-1} \mathcal{G}[\mathsf{a}^{j}] \text{,} \\
        &\mathsf{a}^{j+1} = \tilde{\mathsf{a}}^{j} - \mathbf{P} (\mathcal{M}'[\mathsf{a}^{j}])^\top \left[ (\mathcal{M}'[\mathsf{a}^{j}]) \mathbf{P} (\mathcal{M}'[\mathsf{a}^{j}])^\top + \mathbf{R_a} \right]^{-1} \mathcal{M}[\tilde{\mathsf{a}}^{j}] \text{,}
        \label{eq:REnKF}
    \end{aligned}
\end{equation}
where $j$ is the iteration index, and the prime symbol indicates the corresponding functional gradient with respect to the geometric parameter~$\mathsf{a}$.
The first step is to regularize the geometric parameter based on the total variation of the trailing-edge profile, while the second step aims to update the geometric parameter by minimizing the objectives.
In the formula, the prediction $\mathcal{M}[\tilde{\mathsf{a}}^{j}]$ associated with the regularized geometric parameters~$\tilde{\mathsf{a}}$ is approximated using linear assumption as
\begin{equation}
    \mathcal{M}[\tilde{\mathsf{a}}^{j}] \approx \mathcal{M}[\mathsf{a}^{j}] - \lambda \; \text{cov}(\mathcal{M}[\mathsf{a}^{j}], \mathcal{G}[\mathsf{a}^{j}])\mathbf{W}^{-1} \mathcal{G}[\mathsf{a}^{j}] \text{.}
    \label{eq:Mta}
\end{equation}

The updated sample covariance $\text{cov}(\tilde{\mathsf{a}}^{j},\tilde{\mathsf{a}}^{j})$ in the first step of Eq.~\eqref{eq:REnKF} is scaled relative to the original covariance as
\begin{equation}
    \begin{aligned}
        & \text{cov}(\tilde{\mathsf{a}}^{j},\tilde{\mathsf{a}}^{j}) = \frac{1}{M-1}\left(\tilde{\mathsf{a}}^{j} - \bar{\tilde{\mathsf{a}}}^{j} \right)\left(\tilde{\mathsf{a}}^{j} - \bar{\tilde{\mathsf{a}}}^{j} \right)^\top \text{, with}
        \\
        & \tilde{\mathsf{a}}^{j} - \bar{\tilde{\mathsf{a}}}^{j}  = (\mathsf{a}^{j} - \mathsf{\bar{a}}^{j}) \left(\mathbf{I} - \frac{\lambda}{M-1}(\mathcal{G}[\mathsf{a}^{j}] - \overline{\mathcal{G}[\mathsf{a}^{j}]})^\top \mathbf{W}^{-1} (\mathcal{G}[\mathsf{a}^{j}] - \overline{\mathcal{G}[\mathsf{a}^{j}]})\right) \text{,}
    \end{aligned}
\end{equation}
where $M$ is the ensemble size, the symbol $\bar{\cdot}$ represents the sample mean, and $\mathbf{I}$ is the identity matrix.
The eigenvalues of $\left(\mathbf{I} - \frac{\lambda}{M-1}(\mathcal{G}[\mathsf{a}^{j}] - \overline{\mathcal{G}[\mathsf{a}^{j}]})^\top \mathbf{W}^{-1} (\mathcal{G}[\mathsf{a}^{j}] - \overline{\mathcal{G}[\mathsf{a}^{j}]})\right)$ can exceed 1, especially when the parameter $\lambda$ is large.
Hence the first step in Eq.~\eqref{eq:REnKF} may lead to an increase in the sample covariance, i.e., $\text{cov}(\tilde{\mathsf{a}}^{j},\tilde{\mathsf{a}}^{j})>\text{cov}(\mathsf{a}^{j},\mathsf{a}^{j})$, potentially causing divergence in the optimization process for large regularization parameters~$\lambda$. 
Additionally, Eq.~\eqref{eq:Mta} is given based on linear assumption, which may introduce significant errors in a large regularization step when the model operator~$\mathcal{M}$ is highly nonlinear.
On the contrary, small regularization parameter~$\lambda$ may have negligible effects on the optimization process, resulting in unsmooth optimal trailing edges.
To alleviate these issues, in the conventional regularized ensemble Kalman method~\citep{zhang2020regularized}, a ramp function is used for the regularization parameter~$\lambda$ to gradually increase the strength of the regularization term.
Here, we reformulate the update schemes by augmenting the weight matrix with the sample covariance of $\mathcal{G}[\mathsf{a}]$ as follows
\begin{equation}
    \begin{aligned}
        &\tilde{\mathsf{a}}^{j} =  \mathsf{a}^{j} - \lambda\mathbf{P}(\mathcal{G}^\prime[\mathsf{a}^{j}])^\top \left[\lambda (\mathcal{G}^\prime[\mathsf{a}^{j}])\mathbf{P}(\mathcal{G}^\prime[\mathsf{a}^{j}])^\top + \mathbf{W} \right]^{-1} \mathcal{G}[\mathsf{a}^{j}] \text{,} \\
            &\mathsf{a}^{j+1} = \tilde{\mathsf{a}}^{j} - \mathbf{P} (\mathcal{M}'[\tilde{\mathsf{a}}^{j}])^\top \left[ (\mathcal{M}'[\tilde{\mathsf{a}}^{j}]) \mathbf{P} (\mathcal{M}'[\tilde{\mathsf{a}}^{j}])^\top + \mathbf{R_a} \right]^{-1} \mathcal{M}[\tilde{\mathsf{a}}^{j}] \text{.}
    \end{aligned}
    \label{eq:update_scheme}
\end{equation}
The term~$\lambda (\mathcal{G}^\prime[\mathsf{a}^{j}])\mathbf{P}(\mathcal{G}^\prime[\mathsf{a}^{j}])^\top$ is introduced into the regularization step, which ensures that the update scheme can degrade to the original scheme when $\lambda$ approaches $0$.
Moreover, the sample covariance reduction is guaranteed when $\lambda$ is large~\citep{burgers1998analysis}, which enhances the convergence of the optimization process in the scenario of large shape variations.
The matrices multiplied by $\mathcal{G}[\mathsf{a}]$ and $\mathcal{M}[\mathsf{a}]$ are conventionally regarded as Kalman gain matrices~$\mathbf{K}_r$ and $\mathbf{K}_a$.
The model derivatives in Eq.~\eqref{eq:update_scheme} are estimated using the ensemble-based gradient method, as illustrated in Appendix~\ref{sec:enopt}.

\subsection{Procedure of acoustic shape optimization with ensemble Kalman method}
\begin{figure}[!htb]
    \centering
    \includegraphics[width=1.0\linewidth]{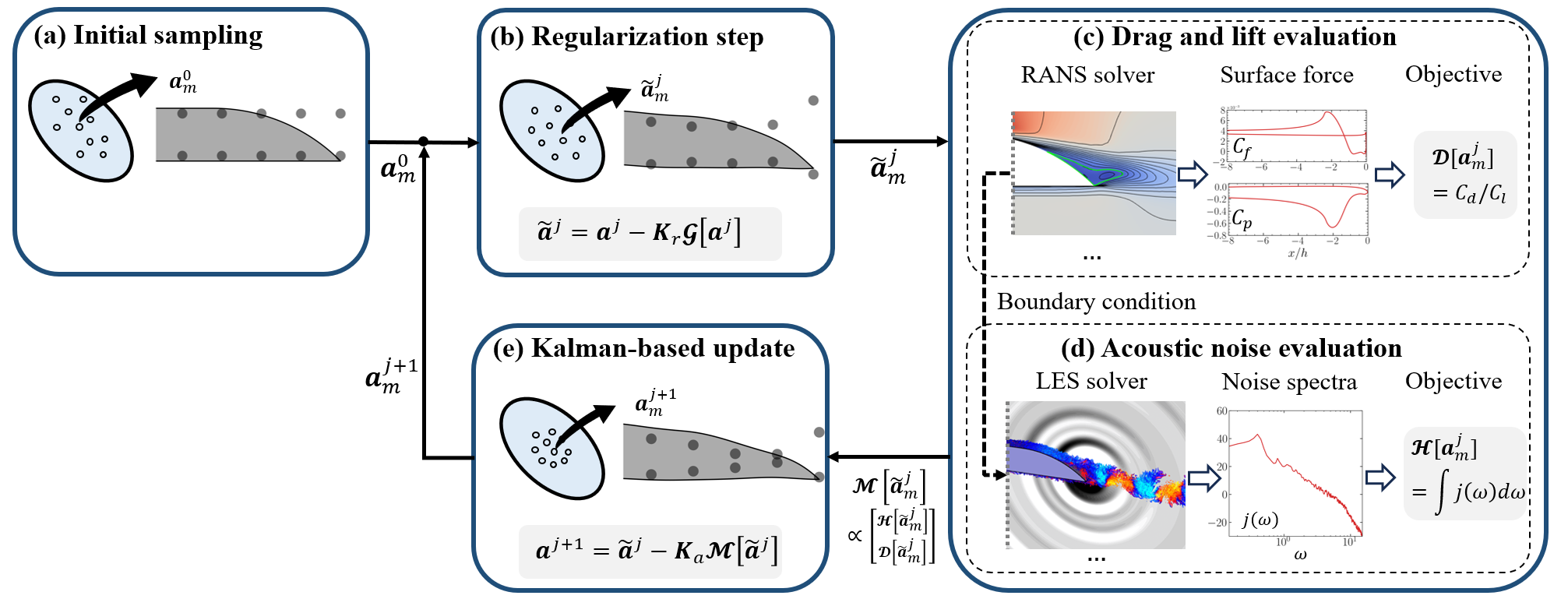}
    \caption{
    Process of the ensemble-based acoustic shape optimization. 
    }
    \label{fig:procedure}
\end{figure}

The procedure of the ensemble Kalman method for the multi-objective optimization process is presented in Figure~\ref{fig:procedure}.
Given the weight parameter $w$, the baseline value~$\mathbf{a}_0$ of control points, and the sample variances~$\sigma^2$, the shape optimization process can be detailed below. 
\begin{enumerate}[label=(\alph*)]
    \item~Initial sampling. 
    The coordinates of all points in the control box are first obtained based on the prescribed control points using the volumetric B-spline interpolation method.
    Further, initial samples~$\{\mathsf{a}_m^0\}_{m=1}^M$ of shape parameter are randomly generated from a normal distribution with the mean value~$\mathsf{a}_0$ and the standard deviation~$\sigma$.\label{process:init_samples}
    \item~Regularization step. The sampled control parameters are regularized to ensure geometric smoothness based on Eq.~\eqref{eq:update_scheme}.
    The regularized control points $\{\tilde{\mathsf{a}}_m^j\}_{m=1}^M $ are subsequently used to deform the trailing-edge shape and corresponding mesh grids for zonal large-eddy simulations.
    \label{process:reg}
    \item~Drag and life evaluation. 
    The RANS-based predictions are performed using the deformed airfoil shapes. 
    This step involves calculating the drag-to-lift ratio $ \{\mathcal{D}[\mathsf{a}_m^j]\}_{m=1}^M $ based on Eq.~\eqref{eq:Da} and generating the inlet flow conditions for the LES around the trailing edge.\label{process:rans}
    \item~Acoustic noise evaluation. The LES-based predictions are performed for the deformed trailing edge to estimate the total acoustic power $ \{\mathcal{H}[\mathsf{a}_m^j]\}_{m=1}^M $ based on Eqs.~\eqref{eq:St}~\eqref{eq:jw}~\eqref{eq:Ha}.
    \label{process:les}
    \item~Kalman-based update. The ensemble-based Kalman scheme is applied to update the position of control points based on Eq.~\eqref{eq:update_scheme}.
\end{enumerate}

The process iterates from Step~\ref{process:reg} until the maximum number of iterations is reached.
To find the Pareto frontier, the weight $w$ is uniformly varied across the range of $[0, 1]$ to explore the trade-offs between the two objectives.
It should be noted that this optimization process involves significant computational effort, primarily due to the multiple iterations of LES simulations. 
For instance, given the maximum iteration number $N$, the sample number $M$, and the scanned number $N_w$ of weight, one needs to conduct $M \times N \times N_w$ large-eddy simulations.
We also note that the ensemble method is inherently parallelizable~\citep{kovachki2019ensemble}, which does not require intercommunication between samples of the LES.
As such, one can improve optimization efficiency by performing multiple LES simultaneously at each update iteration with different weight parameters and sampled geometric parameters.
In this work an open-source platform OpenFOAM~\citep{opencfd21openfoam} is used for zonal large-eddy simulations.
Moreover, the DAFI code~\citep{strofer2021dafi} is used to implement the regularized ensemble-based optimization algorithm.

\section{Optimization results} \label{sec:results}

\subsection{Optimization process}

In this work, the shape optimization problem of the trailing edge in Section~\ref{sec:problem} is solved with an ensemble size~$M=10$ to achieve a balance between the computational efficiency and the optimization accuracy.
For each optimization process, the maximum number of iterations ($N$) is set to 10. Therefore, the total number of LES runs for obtaining the optimal solution is 100.
The initial samples are drawn from a Gaussian distribution with the mean $0$ and the sample variance $0.01$.
The weight \(\mathbf{R}\) and \(\mathbf{Q}\) are set to be \(0.001\) based on our sensitivity test.
A large value leads to a slow convergence speed, while small values would cause an excessive update step and divergence of the optimization process.
The weight \(\mathbf{W}\) is an identity matrix scaled by $0.001$, the rank of which corresponds to the number of grid points defining the trailing-edge shape.

\begin{figure}[!htb]
    \centering
        \centering
        \includegraphics[width=0.8\linewidth]{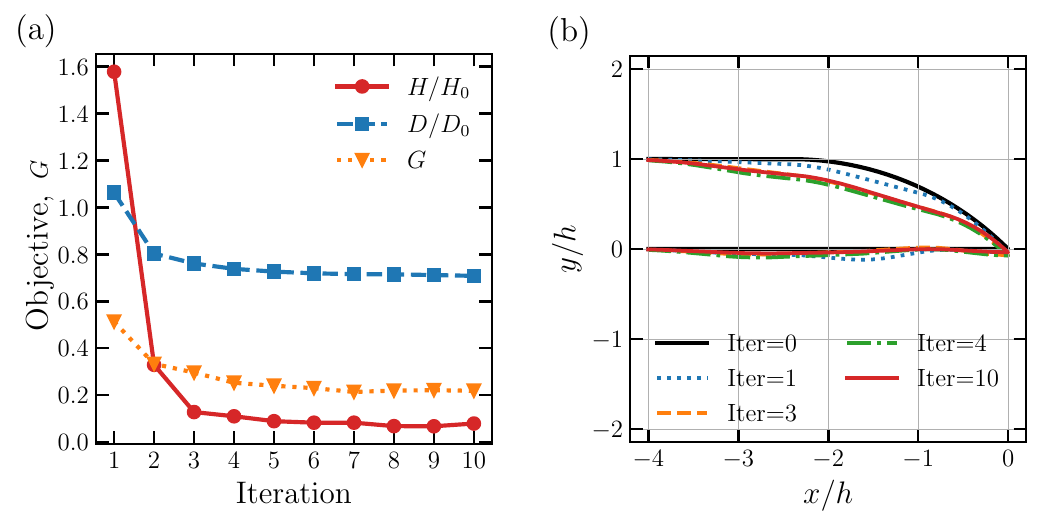}
        \caption{(a) Convergence of the optimization process over ten iterations, in terms of total acoustic power~$H$, drag-to-lift ratio~$D$, and total variation regularization~$G$. 
        (b) The trailing-edge shapes at different iteration steps.
        }
        \label{fig:Loss}
    \label{fig:opt-process}
\end{figure}

Figure~\ref{fig:opt-process} shows the optimization process along the iteration step with the regularization parameter \(\lambda=30\) and the weight parameter $w=0.3$.
It illustrates the convergence behavior of an optimization process over $10$ iterations, in terms of the total acoustic power~$H$, the drag-to-lift ratio~$D$, and the total variation regularization~$G$.
The plotted objective quantities represent the sample-averaged value at each iteration and are normalized by the acoustic power~$H_0$ and the drag-to-lift ratio~$D_0$ of the baseline shape, respectively.
The normalized total acoustic power exhibits a pronounced decrease from their initial values around $1.6$ to $0.078$ at the final iteration.
The total acoustic power of the optimal shape is reduced by $92.2\%$ compared to the baseline shape.
Correspondingly, the far-field SPL is reduced by 10.9 dB.
The initial values exceed $1$ because the noise power calculated with the initial sample mean of shape parameters is larger than that of the baseline shape due to sampling errors.
This also confirms that even small changes in the trailing-edge shape can lead to significant variations in noise power~\citep{Blake2017,Blake2017Vol2}.
The normalized drag-to-lift ratio \(D/D_0\) and the total variation regularization \(G\) both experience a notable decrease from their initial value at the first three iterations and saturate at approximately $0.7$ and $0.2$, respectively.
This indicates that the method can efficiently reduce the cost function within the first few iterations.
The three curves follow a similar decreasing trend and remain nearly unchanged after the fourth iteration step.
This can also be seen from Figure~\ref{fig:opt-process}(b), which shows the shape profiles at different iteration steps.
The geometric shape is almost unvaried after the fourth iteration step, indicating the convergence of the optimization process.

\subsection{Pareto frontier}

The Pareto frontier is obtained by combining the weighted sum method and the ensemble-based optimization method as described in Section~\ref{sec:method}.
The multi-objective optimization problem is solved with a weight increment with an interval of $\Delta w = 0.1$ in this work.
The regularization parameter is set as $\lambda=30$.

\begin{figure}[!htb]
    \centering
    \includegraphics[width=0.9\linewidth]{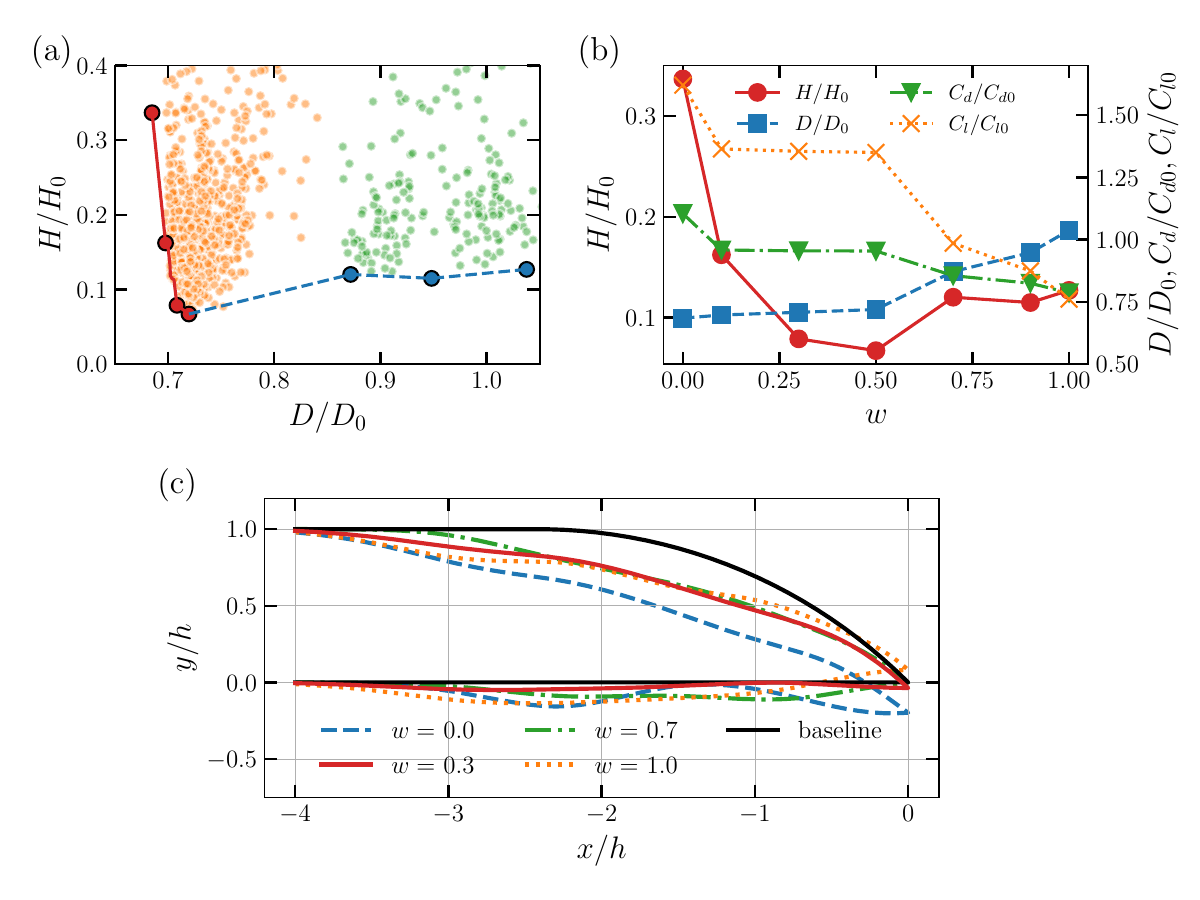}
    \caption{
    Multi-objective shape optimization results with the ensemble Kalman method.
    (a) Pareto frontier. (b) The achieved objective values. (c) The optimized shapes.
    }
    \label{fig:pareto_fontier}
\end{figure}

Figure~\ref{fig:pareto_fontier}(a) shows the obtained Pareto frontier consisting of the optimal solutions based on different weight parameters.
The dots with white edges represent all the sample points generated during the optimization process, defining the feasible region of solutions.
The boundary of the entire feasible region is marked with lines.
The relatively large points represent optimal results obtained with different $w$ values.
The red points from left to right represent the optimality with $w$ of $0, 0.1, 0.3, 0.5$, respectively, and the blue points from left to right indicate the result with $w$ of $0.7, 0.9, 1.0$, respectively.
It can be seen that the entire feasible region can be divided into two parts, represented by the areas occupied by the orange and green points.
This indicates the non-convexity of the objective function, which may cause optimal solutions to fall into local minima, resulting in local Pareto optimality on the blue line.
The red line represents the global Pareto frontier since no point in the feasible region is superior in both objectives to the points on this line.

Figure~\ref{fig:pareto_fontier}(b) presents the objective quantities of the optimized trailing edge, i.e., the total acoustic power and the lift and drag coefficients, varying with the different weight $w$.
As the weight $w$ increases, the total acoustic power~$H$ of the optimized shape generally decreases.
In contrast, the drag-to-lift ratio~$D$ gradually increases as its corresponding weight, i.e., $1-w$, decreases in the cost function.
The total acoustic power decreases in the range from $0$ to $0.5$ monotonically, while getting increased with $w=0.7$ and almost unchanged from 0.7 to 1.0.
That is likely due to the optimization process with $w=0.7$ falling into the local Pareto optimality as shown in Figure~\ref{fig:pareto_fontier}(a).

Figure~\ref{fig:pareto_fontier}(c) presents the optimized trailing-edge shapes obtained for different values of $w$. 
All shapes have a smooth upper surface profile, which reduces flow separation and acoustic noise, as detailed in Section~\ref{sec:Optimal}.
Another noticeable feature is that as the weight $w$ decreases, the trailing edge gradually moves downward, resulting in an increase in the aerodynamic lift as shown in Figure~\ref{fig:pareto_fontier}(b).
The downward movement of the trailing edge increases the angle of the incoming flow.
This can enlarge the pressure difference between the upper and lower surfaces, thereby increasing lift, which is similar to the lift increases with large angles of attack before stall.

The Pareto frontier encapsulates the set of optimal solutions that offer the best possible trade-offs between the two objectives.
Each point on this frontier signifies a compromise between the drag-to-lift ratio and the total sound power.
Based on the criterion~\eqref{eq:criterion}, we identify the Pareto-optimal solution corresponding to a weight coefficient of \(w = 0.3\).
This point can offer a significant reduction in both the drag-to-lift ratio and the total noise power compared to the baseline, which is used in the following tests.

\subsection{Effects of regularization parameter}

\begin{figure}[!htb]
    \centering
    \includegraphics[width=0.9\linewidth]{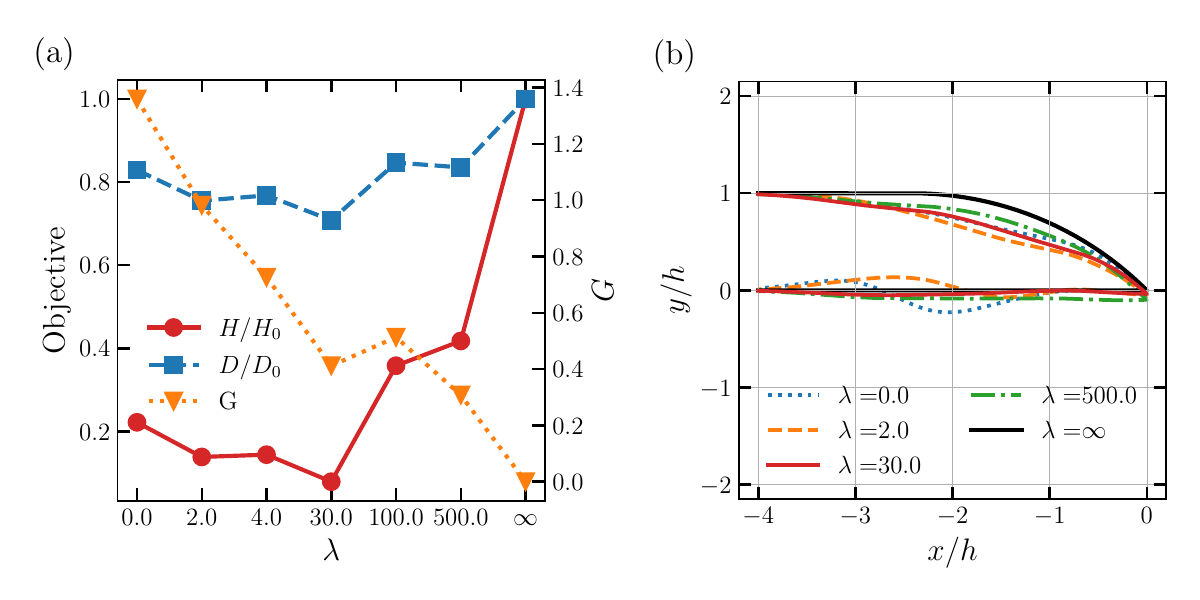}
    \caption{
    Effects of the regularization parameter $\lambda$ on the optimization results, including the relationship between $\lambda$ and quantities of interest ($H/H_0, D/D_0$, and $G$) in the panel (a) and the corresponding optimal shapes in the panel (b). 
    }
    \label{fig:Reg}
\end{figure}

We further examine the effects of the regularization parameter~$\lambda$ on the optimized trailing-edge shape.
Figure~\ref{fig:Reg} illustrates the optimization results with different regularization parameters \(\lambda\).
Figure~\ref{fig:Reg}(a) shows the relationship between the parameter \(\lambda\) and three different quantities of interest, i.e., the total acoustic power~$H$, the drag-to-lift ratio~$D$, and the total variation regularization~$G$.
The values of \(H\) and \(D\) are normalized by the values \(H_0\) and \(D_0\) of the baseline shape, respectively.
In panel (a), the $x$-axis represents the value of $\lambda$, while the left $y$-axis corresponds to the objective value, and the right $y$-axis corresponds to the smoothness regularization. The baseline shape is indicated by the case of $\lambda$ approaching infinity.
As the regularization parameter increases to $\lambda=30$, both optimization objectives decrease.
Further increasing the regularization parameter would deteriorate the acoustic noise and the drag-to-lift ratio of the optimal trailing edge.
With $\lambda=30$, both the acoustic power and drag-to-lift ratio achieve the least value, and the total variation is also significantly reduced compared to the unregularized case of $\lambda=0$.

Figure~\ref{fig:Reg}(b) shows the optimal shapes obtained with different regularization parameters \(\lambda\), where the baseline shape is indicated by the case of \(\lambda=\infty\). 
The regularization parameter \(\lambda\) is crucial in balancing the trade-offs between the objectives and geometric smoothness in the optimization process. 
It can be seen that the optimal shape becomes smoother with $\lambda$ increasing due to the enhanced influence of the total variation regularization term.
However, an excessively large regularization parameter, e.g., $\lambda=500$, can constrain the shape variations, limiting the optimal solution in the vicinity of the baseline shape. 
On the contrary, a small regularization parameter, e.g., $\lambda=2$, would lead to an unsmooth trailing-edge shape.
Given this, we choose the regularization parameter $\lambda=30$ to achieve the significant reduction of objective values with a smooth trailing-edge shape based on our numerical tests.

\section{Optimal trailing edges} \label{sec:Optimal}

\subsection{Time-averaged flow statistics}

\begin{figure}[!htb]
    \centering
    \includegraphics[width=0.95\linewidth]{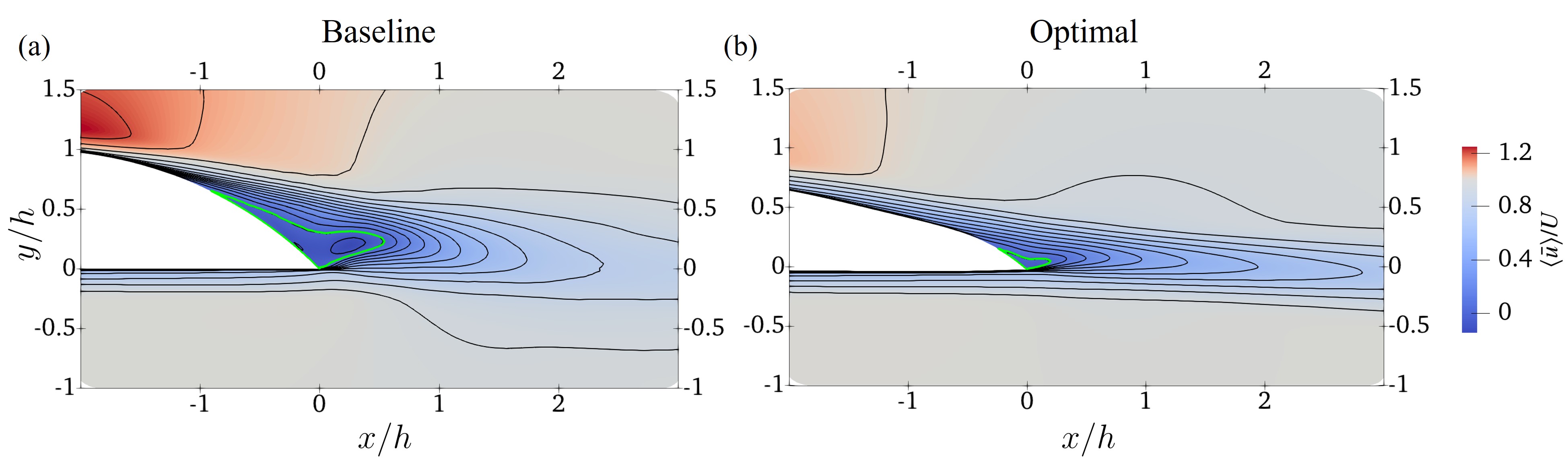}
    \caption{Comparison between (a) the baseline and (b) the optimal shapes in the mean streamwise velocity field $\langle \bar{u} \rangle/U$ computed from the LES. 
    The level of 0 is highlighted with a green line. }
    \label{fig:UMean-Optd}
\end{figure}

To better understand the aerodynamic and acoustic improvements, it is necessary to examine the time-averaged flow statistics around the optimized trailing edge.
Figure~\ref{fig:UMean-Optd} presents a comparison between the baseline and the optimal shapes in the mean streamwise flow field, denoted as $\langle \bar{u}\rangle /U$.
The flow field is obtained through the LES by averaging over the homogeneous spanwise direction and time.
The mean velocity field of the baseline shape in Figure~\ref{fig:UMean-Optd}(a) exhibits a significant separation region behind the trailing edge, characterized by a large area of low velocity.
In contrast, the optimal shape remarkably reduces the areas of the separation region as presented in Figure~\ref{fig:UMean-Optd}(b).
Conclusively, the optimal trailing-edge shape can suppress flow separation and enhance flow attachment to the airfoil surface.

\begin{figure}[!htb]
    \centering
    \includegraphics[width=0.9\linewidth]{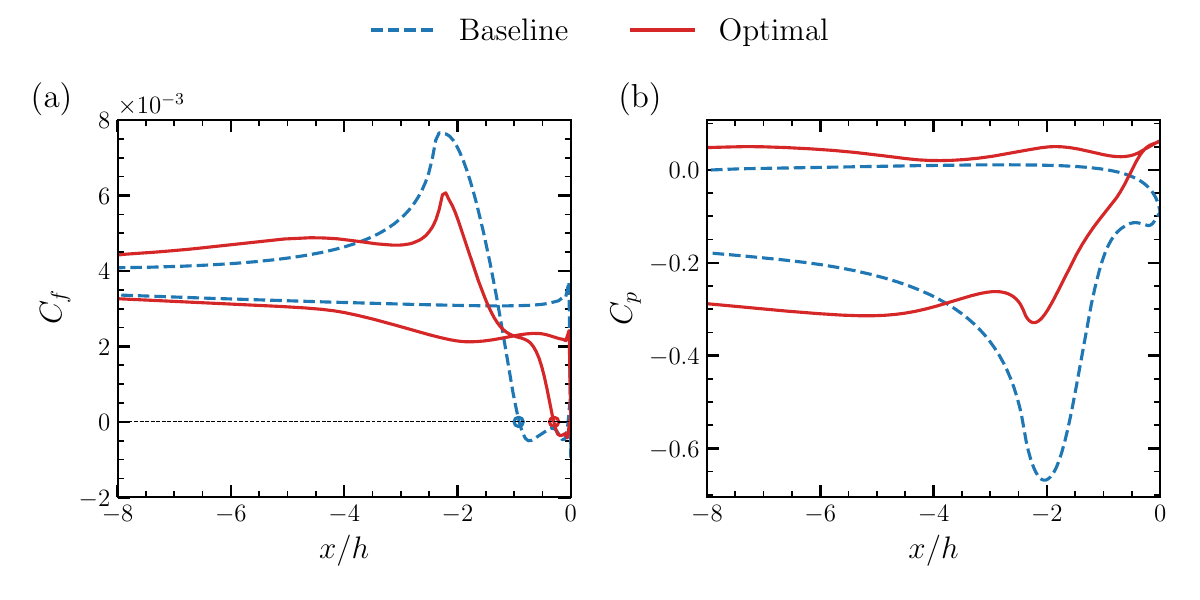}
    \caption{Comparison between the baseline and the optimal shapes in (a) friction coefficient~$C_f$ and (b) pressure coefficient~$C_p$. The circles in the panel (a) indicate the separation locations.}
    \label{fig:CfCp}
\end{figure}

The optimal trailing-edge shape can significantly reduce the drag force and increase the lift force by suppressing the flow separation.
Specifically, the drag coefficient decreases by 5.6\% from 0.0267 to 0.0252, and the lift coefficient increases by 32.9\% from 0.58 to 0.771, compared to the baseline shape.
It is also supported by Figure~\ref{fig:CfCp}, which plots the distribution of friction and pressure coefficients of the baseline and optimal shapes.
The separation points are marked in Figure~\ref{fig:CfCp}(a) with the zero-value point of $C_f$, showing that the separation points are delayed noticeably.
Moreover, it can be seen that the friction coefficients of the lower and the upper walls decrease overall, while that of the upper wall increases from $x/h=-8$ to $-4$.
This is likely due to the reduced pressure gradient allowing large normal gradients of the streamwise velocity and further the friction within the boundary layer~\citep{thompson1985characteristics}.
Figure~\ref{fig:CfCp}(b) shows that the adverse pressure gradient at the beveled position, i.e., ranging from $x/h=-2$ to $0$, is significantly reduced with the optimal shape, which leads to a more favorable pressure distribution and further pressure drag reduction.
Besides, the larger pressure difference between the upper and lower walls can be observed, which results in an increase in aerodynamic lift.

\begin{figure}[!htb]
    \centering
    \includegraphics[width=0.95\linewidth]{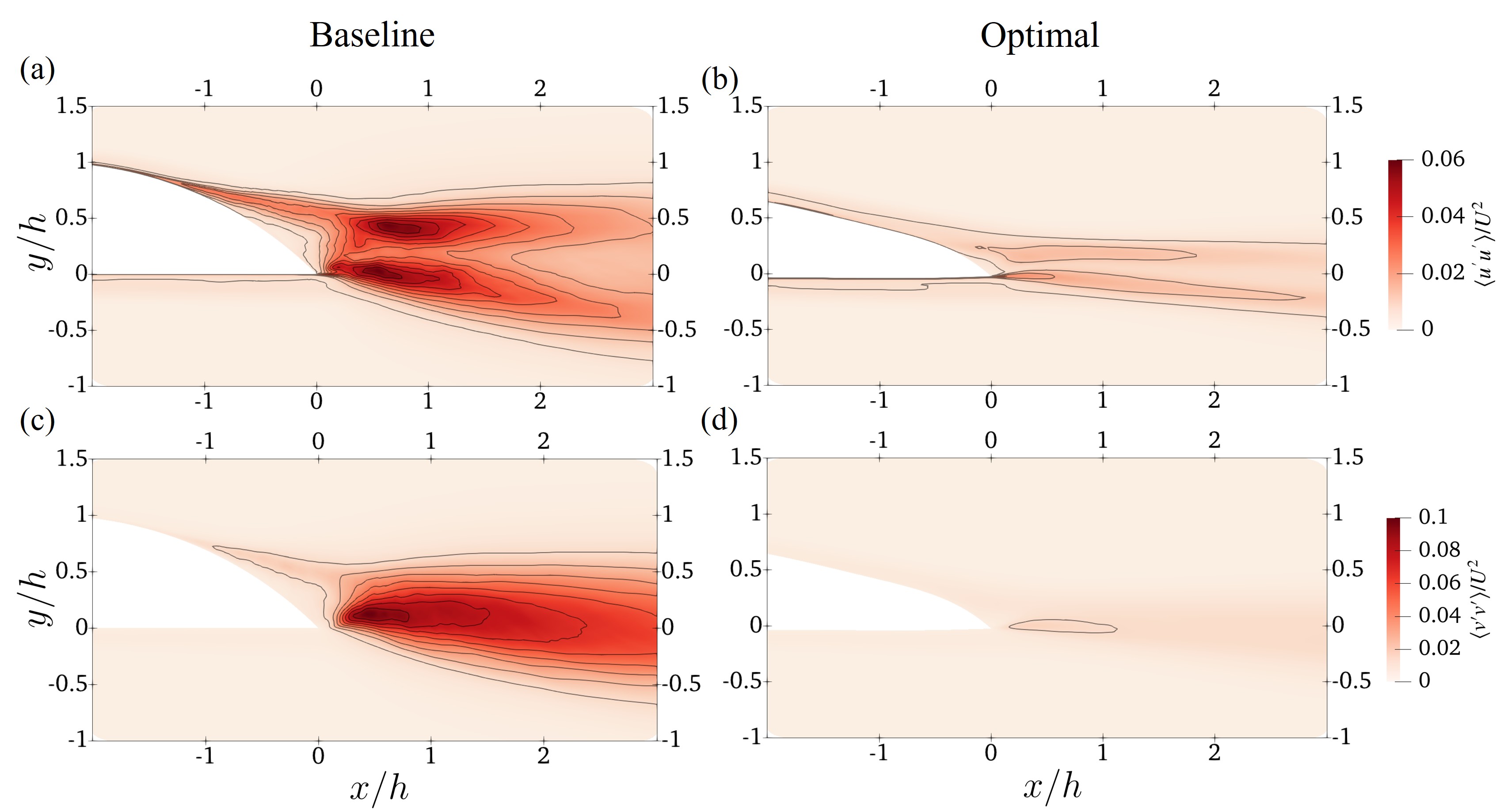}
    \caption{Comparison between the baseline and the optimal shapes in the Reynolds normal stress~$\langle u^\prime u^\prime \rangle/U^2, \langle v^\prime v^\prime \rangle/U^2$ computed from LES.}
    \label{fig:u2-Optd}
\end{figure}

The velocity fluctuations are significantly suppressed with the optimal trailing edge, which leads to the acoustic power reduction.
It is evident in Figure~\ref{fig:u2-Optd} that shows the normalized Reynolds normal stress components of the baseline and optimal shapes. 
It can be seen that the regions with strong turbulence intensity are located at the shear layers in the wake, as indicated by the two peaks in the component $\langle u^\prime u^\prime \rangle$.
Moreover, the turbulence intensity of the optimal shape is significantly lower than the baseline shape.
The overall reduction of turbulence intensity especially in the component~$\langle v^\prime v^\prime \rangle$ can indicate the trailing edge noise reduction because the velocity fluctuations normal to the trailing edge provide the predominant acoustic source~\citep{williams1970aerodynamic}.

\subsection{Far-field noise}
\begin{figure}[!htb]
        \centering
        \includegraphics[width=0.9\linewidth]{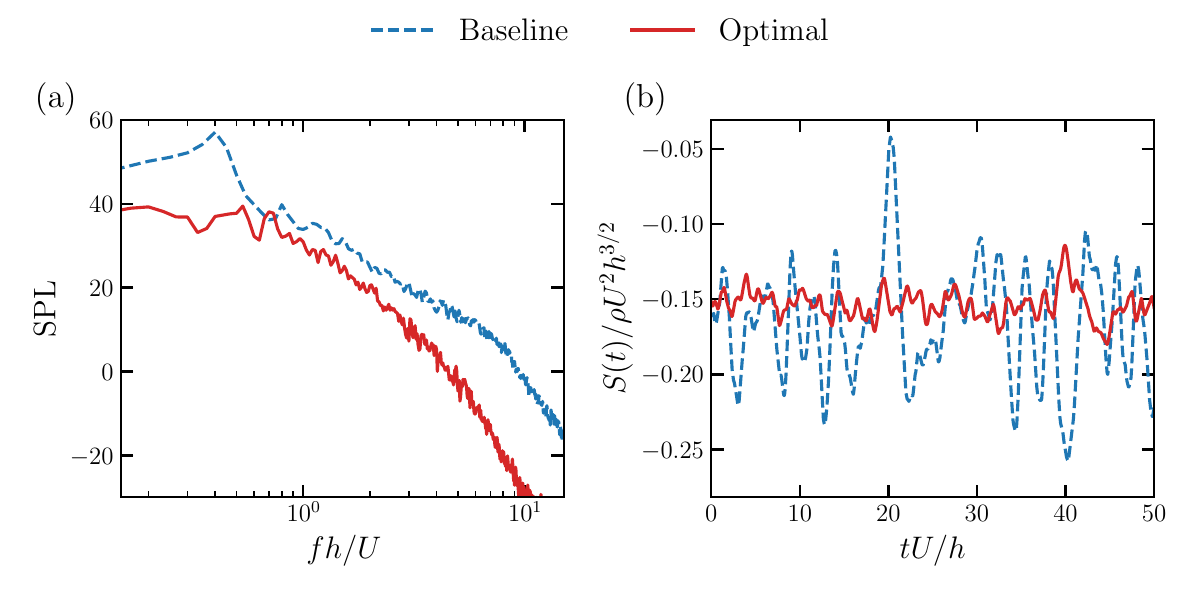}
        \caption{Comparison between the baseline and the optimal shapes in (a) the far-field SPL spectrum and (b) the noise source term~$S(t)$ in Eq.~\ref{eq:St}. 
        }
        \label{fig:SPL_optd}
\end{figure}

The optimal trailing-edge shape can significantly reduce the far-field noise compared to the baseline shape.
It can be observed in Figure~\ref{fig:SPL_optd}(a) which presents the sound pressure level (SPL) of the baseline and the optimal shapes at the non-dimensional frequency~$fh/U$.
The optimal shape exhibits a significant reduction in the SPL, particularly in the low and high-frequency range.
Also, we present that the noise source term~$S(t)$ from Eq.~\eqref{eq:St} changes over time in Figure~\ref{fig:SPL_optd}(b). 
It shows that the amplitude of the noise source term~$S(t)$ is remarkably decreased with the optimal shape, compared to the baseline trailing edge.
The reduction in low-frequency noise is likely due to the reduced bevel angle suppressing large-scale vortex shedding, while the reduction in high-frequency noise can be caused by the decrease in the turbulence length scale, which will be discussed in Section~\ref{sec:mechanism}.
Our optimization results demonstrate that a thinner trailing edge effectively leads to reduced drag-to-lift ratio and suppressed noise emission.
This finding is consistent with the literature~\cite{Kang2025Effect} that the thin trailing edge can enhance both the aerodynamic and aeroacoustic performance.

\subsection{Velocity spectrum}

\begin{figure}[!htb]
    \centering
    \includegraphics[width=.9\linewidth]{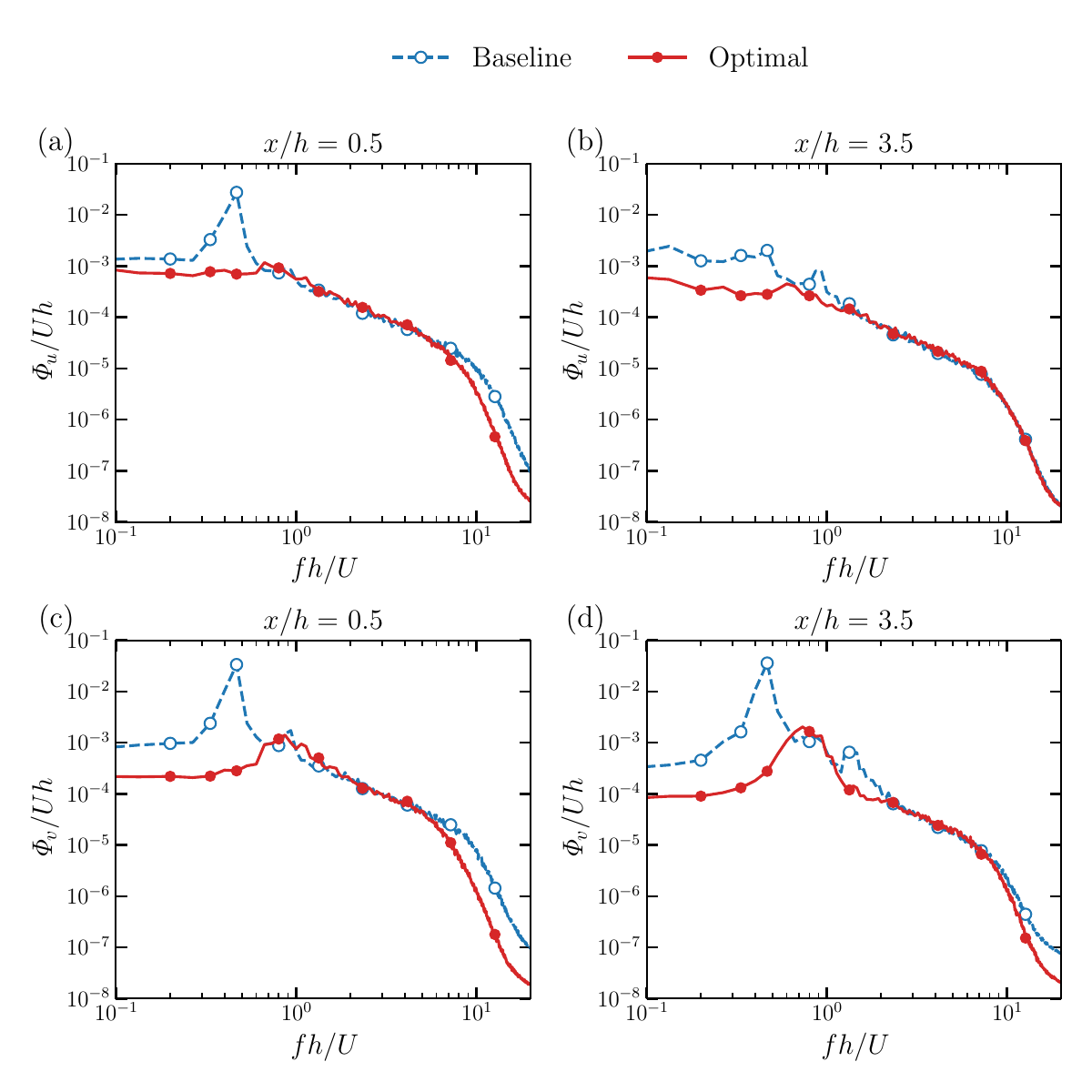}
    \caption{Comparison between the baseline and the optimal shapes in the velocity power spectral density at the two locations, i.e., $x/h=0.5, 3.5$ and $y/h=0$. 
    }
    \label{fig:velocty_psd}
\end{figure}

We present the velocity power spectral density~(PSD) of the baseline and optimal shapes to show the velocity characteristics in the frequency domain. 
Figure~\ref{fig:velocty_psd} shows a comparison of the velocity PSD between the baseline and the optimal shapes at two locations downstream, i.e., $(x/h, y/h)=(0.5,0), (3.5,0)$.
At the near-wake position $x/h = 0.5$, both the streamwise and normal velocity spectra have noticeable peaks at the frequency of $fU/h = 0.4$, consistent with the experimental observation~\citep{olson2004experimental}. 
At the relatively far-wake position $x/h = 3.5$, the streamwise and normal velocity fluctuations, which are defined as $u_i^{\text{rms}} = \int_0^\infty \Phi_{u_i}(\omega) d\omega$, is mitigated compared to the upstream due to energy dissipation. 
Meanwhile, the spectral peak of the streamwise velocity shows a marked decrease in the far wake, while that of the normal velocity remains almost unchanged. 
This is likely attributed to the redistributive effect of the fluctuating pressure, transferring energy from $\langle {u^\prime}^2 \rangle$ to $\langle {v^\prime}^2 \rangle$~\citep{pope2001turbulent,Bastankhah2024}.
The optimal shape leads to a significant reduction in the velocity spectrum in the low-frequency range. 
Particularly, the first peak observed in the baseline shape is suppressed with the optimal shape.
Moreover, the integration of velocity PSD can indicate the reduction in root-mean-square (RMS) of the velocity fluctuations.
This suppression in velocity fluctuations, particularly in the low-frequency range, can result in the trailing edge noise reduction based on Eq.~\eqref{eq:pa}, which will be further discussed in Section~\ref{sec:lighthill}.

\begin{figure}[!htb]
    \centering
    \includegraphics[width=1.0\linewidth]{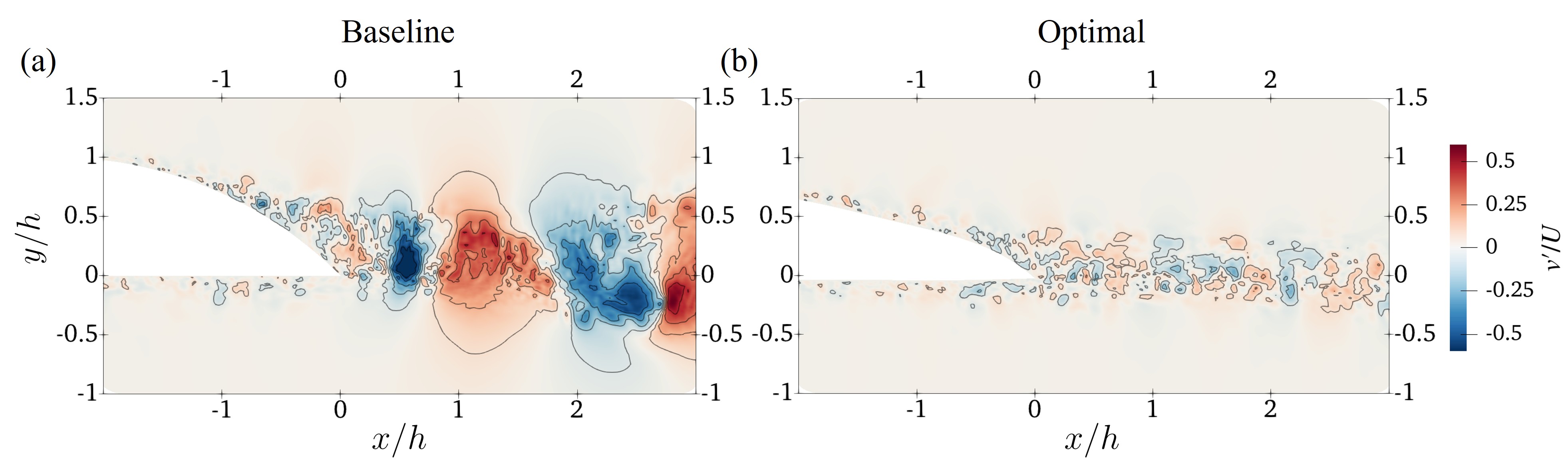}
    \caption{Comparison between (a) the baseline shape and (b) the optimal shape in the instantaneous normal velocity fluctuations $v^\prime/U$ computed from the LES.
    }
    \label{fig:u}
\end{figure}

The suppression of the first peak in the velocity spectrum indicates that the optimal shape effectively mitigates the vortex shedding.
It is supported in Figure~\ref{fig:u}, showing the instantaneous normal velocity fluctuations of the baseline and optimal shapes, respectively.
The velocity fluctuations are obtained by subtracting the mean velocity from the instantaneous velocity.
Figure~\ref{fig:u}(a) shows the significant vortex shedding of the baseline shape, characterized by the large amplitude of wake meandering.
In contrast, the optimal shape in Figure~\ref{fig:u}(b) shows a remarkable reduction in the normal velocity fluctuations and the wake meandering.
This indicates that the optimal shape effectively disrupts the formation and shedding of large flow vortices, thereby reducing noise generation~\citep{howe2003theory}.

\subsection{SPOD analysis}
\label{sec:mechanism}

The spectral proper orthogonal decomposition (SPOD) method~\citep{sieber2016spectral,schmidt2018spectral} is employed to further investigate the large-scale flow structures and their impact on the reduction of acoustic noise and drag-to-lift ratio.
The SPOD method is applied to the normal velocity over approximately 14 flow-through times~($14L/U$), which can provide the dominant flow features and their energy distribution over frequency~\cite{li2022onset}.
By analyzing the SPOD modes and their corresponding energy spectra, we aim to show the frequency and structural characteristics of flow fields around the optimal trailing-edge shape.

\begin{figure}[!htb]
    \centering
    \includegraphics[width=.9\linewidth]{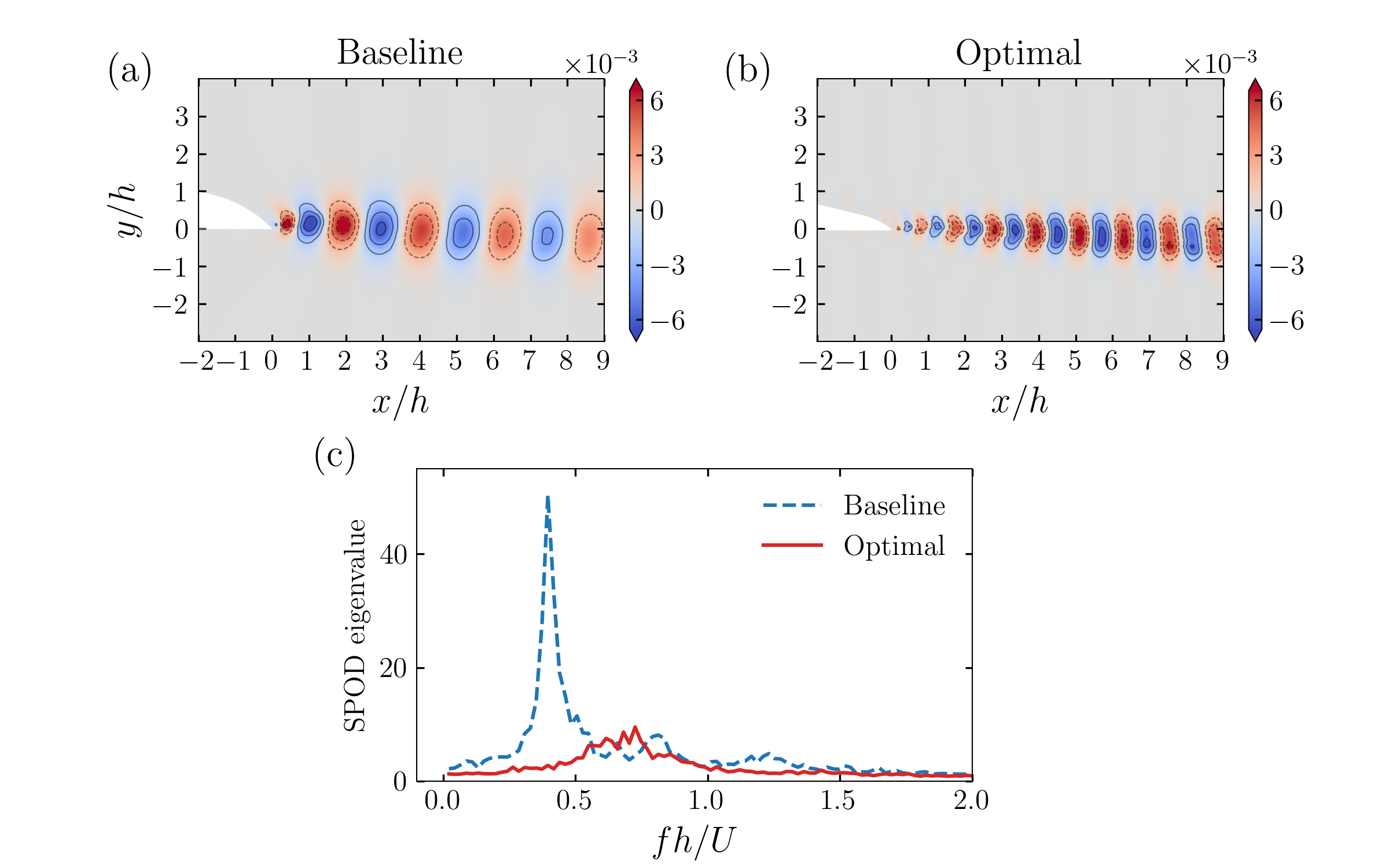}
    \caption{
    Comparison of SPOD modes of the normal velocity between the baseline and optimal shapes. 
    (a, b) First spatial modes of normal velocity at dominant frequencies.
    (c) Maximum SPOD eigenvalues.
    }
    \label{fig:SPOD}
\end{figure}

Figures~\ref{fig:SPOD}(a) and (b) present the comparison of the leading SPOD modes of the normal velocity at the dominant frequencies between the baseline and optimal shapes, respectively.
The dominant frequency with the highest energy for the baseline shape is $fh/U=0.40$, while that for the optimal shape is $fh/U=0.73$.
Figures~\ref{fig:SPOD}(a) and (b) show that the SPOD modes for the baseline shape have relatively larger amplitudes in the near wake than the optimal shape. 
Moreover, with the optimal shape, the amplitudes of the modes increase along the streamwise direction and gradually decrease for $x/h > 4.5$, while the peak decay occurs at $x/h=0.38$ for the baseline shape.
Hence, the dominant mode with the maximum amplitude is spatially farther away from the trailing edge for the optimal shape.
This can also indicate the noise reduction according to Eq.~\eqref{eq:St}, as the corresponding location $r_0$ associated with significant velocity fluctuation
is increased with the optimal shape~\citep{GLEGG2017365}.
The structures of the dominant spatial modes with the optimal shape are noticeably smaller than those of the baseline shape in both the streamwise and normal directions.
This signifies that the optimal shape reduces the length scale of coherent structures.
Figure~\ref{fig:SPOD}(c) shows the comparison of the maximum SPOD eigenvalues across frequencies for both trailing-edge shapes. 
The peak of the SPOD eigenvalue for the flow field around the optimal shape is significantly reduced compared to the baseline shape. 
The suppression in the eigenvalue peak represents a decrease in the energy associated with the dominant flow structures,
which contributes to the reduction of trailing edge noise.

The length scale of the turbulent flow can affect the trailing edge noise in the high-frequency range.
Specifically, based on the Eq.~\eqref{eq:pa}, the estimated magnitude of high-frequency pressure spectrum can be derived as~\citep{GLEGG2017365}
\begin{equation}
    S_{pp}(\mathbf{x},\omega) \sim  \frac{ \sin{\phi} \sin^2{\left(\frac{1}{2}\theta\right)}\rho^2 U^3 u_s^2 L_s b}{| \mathbf{x} |^2 c_{\infty}}I(\omega) \text{,}
    \label{eq:Spp_high_freq}
\end{equation}
where $u_s$ represents the velocity scale of the turbulent fluctuations, $L_s$ denotes the length scale of turbulence, $b$ is the span of the trailing edge, and $I(\omega)$ provides the shape of the normalized pressure spectrum.
Eq.~\eqref{eq:Spp_high_freq} reveals that the trailing edge noise is proportional to the length scale of turbulence at high frequencies.
Both Figure~\ref{fig:u} and Figure~\ref{fig:SPOD} indicate that the length scale of turbulence is reduced with the optimal trailing edge, which can lead to high-frequency noise reduction according to Eq.~\eqref{eq:Spp_high_freq}.

\subsection{Spectrum of Lighthill stress}
\label{sec:lighthill}

The reduction of the low-frequency velocity spectrum observed in Figure~\ref{fig:velocty_psd} can mitigate the trailing edge noise in both the low-frequency and high-frequency ranges.
This can be examined by investigating the spectrum of the Lighthill stress.
In this work, the viscous and nonisentropic terms are neglected in the Lighthill stress due to the high Reynolds number and low Mach number.
Also, the density fluctuation contribution can be ignored in the Lighthill stress given the low Mach number, leaving only the squared velocity term.
Further, the power spectrum of far-field sound pressure is proportional to the power spectrum of the squared velocity based on Eqs.~\eqref{eq:St} and~\eqref{eq:Phi_pa}.
Given this, the Lighthill stress stems from the nonlinear interaction of velocity fluctuations.
The high-frequency Lighthill stress can be influenced by the low-frequency velocity fluctuations due to the nonlinear nature.
To be specific, the spectrum of the squared velocity can be obtained by the convolution of the velocity spectrum as~\citep{pope2001turbulent}
\begin{align}
    \Phi_{u_iu_j}(\omega) = \frac{1}{2 \pi}\int \Phi_{u_i}(\omega^\prime)\Phi_{u_j}(\omega-\omega^\prime) d\omega^\prime \text{.}
    \label{eq:convol}
\end{align}
This indicates the spectrum of squared velocity at frequency $\omega$ is contributed by velocity spectra at all possible pairs of interacting frequencies $\omega'$ and $\omega-\omega'$.
Particularly, two low-frequency components of velocity fluctuations can synthesize a high-frequency component of Lighthill stress through the convolution in Eq.~\eqref{eq:convol}.
As such, there can be a substantial reduction in the squared velocity spectrum across a wide range, due to the reduced velocity spectrum in low frequencies as shown in Fig.~\ref{fig:velocty_psd}.
Moreover, the far-field sound pressure is proportional to the near-field integral of the Lighthill stress tensor, according to the Lighthill equation~\eqref{eq:pa} and~\eqref{eq:St}.
Therefore, suppressing velocity fluctuation in low frequencies is able to reduce the high-frequency noise~\citep{sandham2006nonlinear}.

\begin{figure}[!htb]
    \centering
    \includegraphics[width=0.9\linewidth]{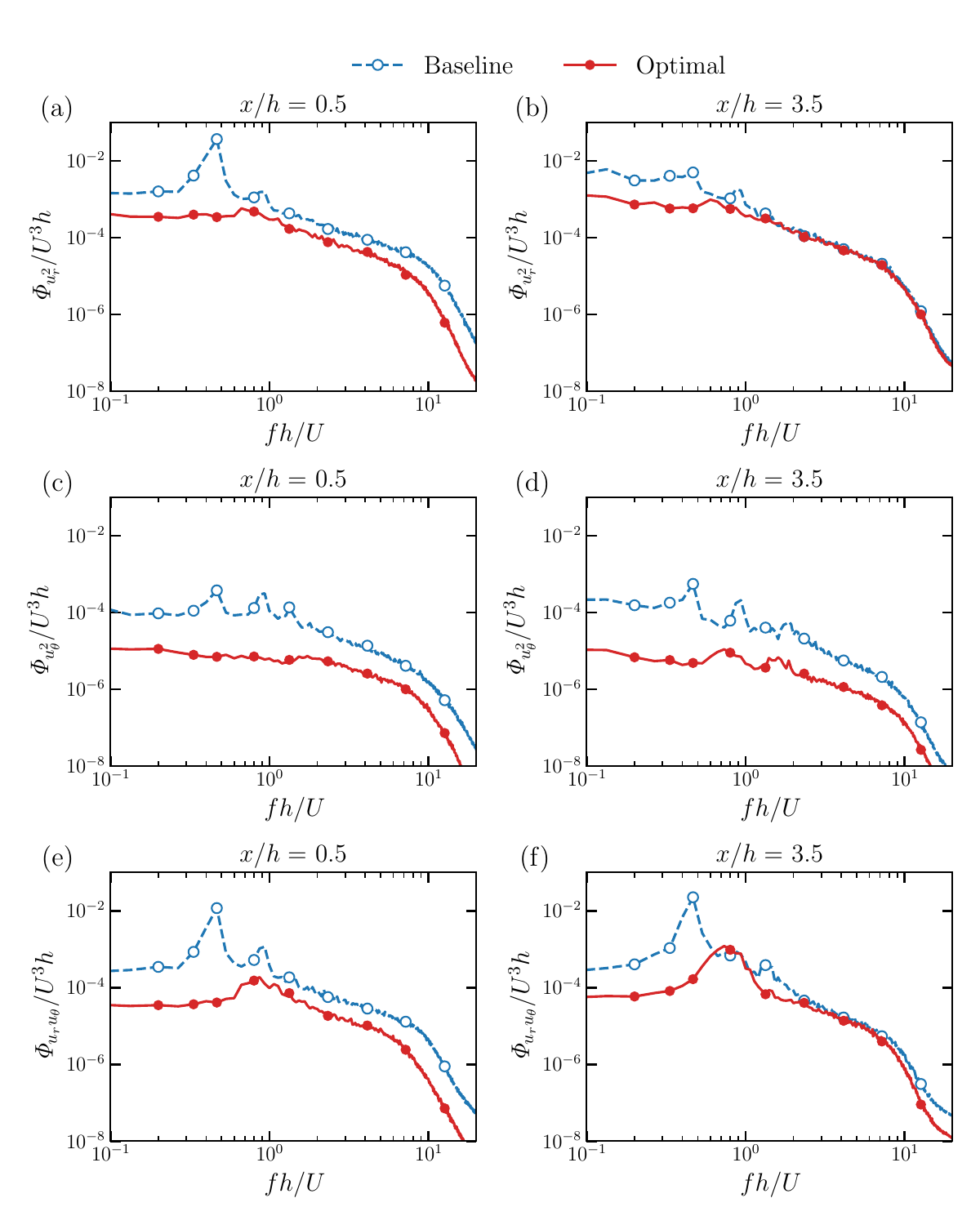}
    \caption{Comparison between the baseline and the optimal shapes in the power spectrum of the squared velocities i.e., $u_r^2$, $u_{\theta}^2$, and $u_r u_\theta$, at the location $x/h=0.5,3.5$ and $y/h=0$. 
    }
    \label{fig:velocty_square_psd}
\end{figure}

Figure~\ref{fig:velocty_square_psd} presents a comparison of the power spectrum of the squared velocities~($u_r^2, u_{\theta}^2,u_ru_\theta$) between the baseline and the optimal shapes at the locations $(x/h,y/h)=(0.5,0),(3.5,0)$.
Notably, the spectrum magnitude of each term in Eq.~\eqref{eq:St} is considerably diminished across a broad frequency range for the optimal shape in contrast to the baseline shape.
This reduction is consistently observed at different downstream locations.
Specifically, at the location~$x/h=0.5$, the optimal shape provides a marked decrease in spectrum magnitude in both the low and high-frequency ranges as shown in Figure~\ref{fig:velocty_square_psd}.
At the location $x/h=3.5$, the spectrum magnitudes of $u_r^2$ and $u_ru_\theta$ are mainly suppressed in the low-frequency range with the optimal trailing edge, while that of $u_\theta^2$ is reduced over the entire frequency range.
The negligible variation of $u_r^2$ and $u_ru_\theta$ in the high frequency is due to the effects of the recovery of mean velocity~$\langle u_r \rangle$ and the attenuation of the fluctuating velocity~$u_r^\prime$ at the far wake.
The power spectrum of squared velocities for the optimal shape indicates that the reduction of noise intensity spans across both low and high frequencies, enhancing the overall acoustic performance of the trailing edge.

\section{Conclusion} \label{sec:conclusion}

This work presents the ensemble Kalman method for reducing the trailing edge noise by optimizing the geometric shape.
A multi-objective function associated with both the acoustic noise and drag-to-lift ratio is formulated to find optimal trailing edges with improved acoustic and aerodynamic performance.
The ensemble Kalman method is developed to find the Pareto-optimal solutions of the multi-objective shape optimization problem.
Moreover, the smoothness regularization step is introduced to avoid unsmooth trailing edges.
In addition, the zonal large-eddy simulation, coupled with the acoustic analogy of the Ffowcs Williams and Hall, is used to evaluate the trailing edge noise, in order to improve the optimization efficiency.

The capability of the present method is demonstrated in a slender airfoil with beveled trailing edges.
Our results demonstrate that the method can efficiently find the optimal trailing edge that reduces the acoustic noise and drag-to-lift ratio simultaneously.
The trailing edge shape is optimized to be more streamlined with a reduced bevel angle, which suppresses the boundary layer separation and the vortex shedding.
The SPOD method is used to analyze the spatial flow characteristics at dominant frequencies.
The results indicate that the optimal shape leads to noise suppression by disrupting large-scale flow structures near the trailing edge.
On the other hand, the Lighthill stress is analyzed, which reveals that the optimal shape significantly suppresses the noise at the high-frequency range due to the nonlinear interaction of reduced low-frequency velocity fluctuations.

Future works will be conducted to explore high-dimensional shape optimizations with various design variables, demonstrating the merits of the ensemble Kalman method in optimization efficiency.
Moreover, the compressible LES coupled with the acoustic analogy of Ffowcs Williams \& Hawkings will be investigated to take into account the effects of scattering directivity in acoustic shape optimization.
Besides, the method can be extended to machine-learning-assisted LES modeling of subgrid-scale stress~\citep{lozano2023machine} and wall shear stress~\citep{zhou2025wall} based on multiple data sources from different flow scenarios.

\appendix

\section{Comparison with the optimal shape using the MADS method}
\label{sec:validate}

\begin{figure}[!htb]
    \centering
    \includegraphics[width=1.0\linewidth]{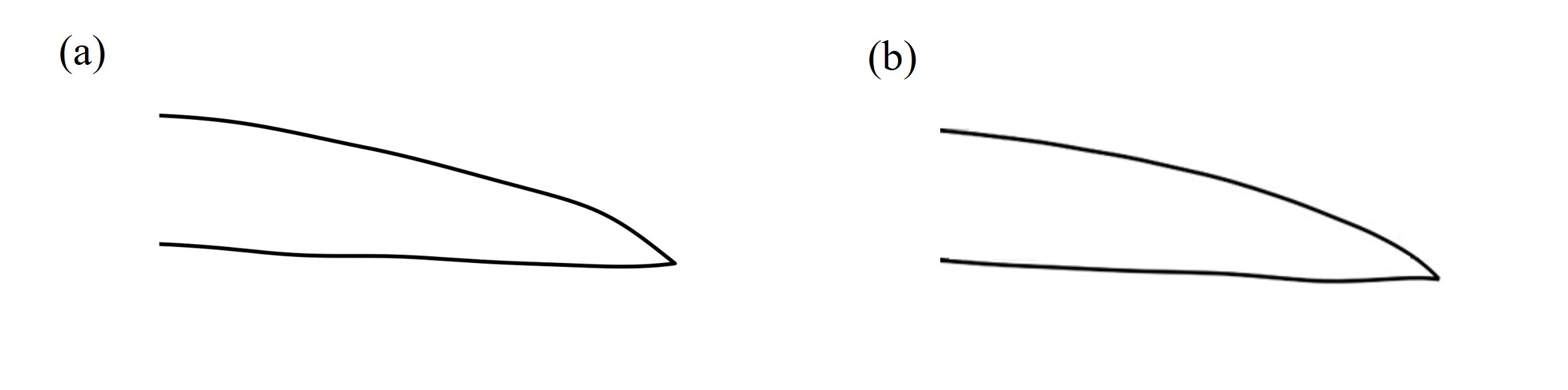}
    \caption{Comparison with the optimal shape with the MADS method. (a) The optimized shape with the ensemble method using the parameters of $w=0.5$ and $\lambda=10$. (b) The optimal shape with the MADS method from~\citet{marsden2007trailing}.}
    \label{fig:compare_MADS}
\end{figure}

The feasibility of the ensemble Kalman method is validated by comparing with the optimal shape obtained with the mesh adaptive direct search (MADS) method from~\citet{marsden2007trailing}.
Figure~\ref{fig:compare_MADS} presents the comparison of trailing-edge shapes optimized using the present method and the MADS method. Figure~\ref{fig:compare_MADS}(a) shows the optimized shape obtained with the present method with the parameters $w=0.5$ and $\lambda=10$, while Figure~\ref{fig:compare_MADS}(b) displays the optimal shape from the MADS method adapted from the work of~\citet{marsden2007trailing}. 
The noise reduction achieved by the obtained shape in Figure~\ref{fig:compare_MADS}(a) is $86.9\%$, which is similar to $89\%$ for the MADS-optimized shape~\citet{marsden2007trailing}.
The results show that our method can achieve an optimal shape similar to that of the MADS method with specific weight and regularization parameters, validating the effectiveness of our optimization method.
Moreover, the chosen parameters in this work, i.e., $w=0.3$ and $\lambda=30$, can lead to an optimal trailing edge that achieves better noise reduction by $92.2\%$.
The improvement may be attributed to the increased number of control points, i.e., 10 control points, compared to the 5 control points of the reference work.
This allows for more flexible shape deformation to enable growing search space for optimal solutions. 

\section{Ensemble-based gradient for LES-based optimization}
\label{sec:enopt}

The gradient of the cost function is required to guide the updated direction of the geometric parameters.
Due to the chaotic effects of turbulence, the adjoint-based method is faced with the blowup issue, i.e., the divergence to infinity, in the gradient calculation.
Here the ensemble-based method is used to estimate the gradient of the physical quantities, e.g., the trailing edge noise, with respect to the shape changes.
Specifically, the samples of shape parameters are $\mathsf{a}_m = \bar{\mathsf{a}} + \varepsilon_m, m=1,2,...,M$, where $M$ is the total number of samples, and $\varepsilon_m$ are Gaussian distributed random perturbations with zero mean and variance $\sigma^2$. 
The physical quantity, such as the total acoustic power at each sample $\mathcal{H}[\mathsf{a}_m]$, can be approximated using a Taylor series expansion around the mean shape parameter $\bar{\mathsf{a}}$ as
\begin{equation}
    \mathcal{H}[\mathsf{a}_m] - \mathcal{H}[\bar{\mathsf{a}}]= \mathcal{H}^\prime \left( \mathsf{a}_m - \bar{\mathsf{a}} \right) \text{.} \label{eq:Taylor_expansion}
\end{equation}
The model gradient $(\mathcal{H}^\prime)^\top$ is often pre-multiplied by the covariance matrix $\mathbf{P} \equiv \text{cov}(\mathsf{a},\mathsf{a}) = \frac{1}{M - 1} \sum_{m=1}^{M}\left( \mathsf{a}_m - \bar{\mathsf{a}} \right)\left( \mathsf{a}_m - \bar{\mathsf{a}} \right)^\top$.
This is equivalent to the sample covariance between the geometric parameter $\mathsf{a}$ and the model prediction $\mathcal{H}[\mathsf{a}]$.
That is,
\begin{equation}
    \mathbf{P}(\mathcal{H}^\prime)^\top = \text{cov}(\mathsf{a},\mathcal{H}[\mathsf{a}]) = \frac{1}{M-1}\sum_{m=1}^{M}\left( \mathsf{a}_m - \bar{\mathsf{a}} \right)\left( \mathcal{H}[\mathsf{a}_m] - \mathcal{H}[\bar{\mathsf{a}}] \right)^\top \text{,}
    \label{eq:PHT}
\end{equation}
where the physical quantity at the mean is often approximated by the sample mean, i.e., $\mathcal{H}[\bar{\mathsf{a}}] \approx \overline{\mathcal{H}[\mathsf{a}]}$.
This approach can be also used to estimate $(\mathcal{H}^\prime)\mathbf{P}(\mathcal{H}^\prime)^\top$ as
\begin{equation}
    (\mathcal{H}^\prime)\mathbf{P}(\mathcal{H}^\prime)^\top = \text{cov}(\mathcal{H}[\mathsf{a}],\mathcal{H}[\mathsf{a}]) \text{,}
    \label{eq:HPHT}
\end{equation}
with the linear approximation in Eq.~\eqref{eq:Taylor_expansion}.

\begin{figure}[!htb]
    \centering
    \includegraphics[width=1.0\linewidth]{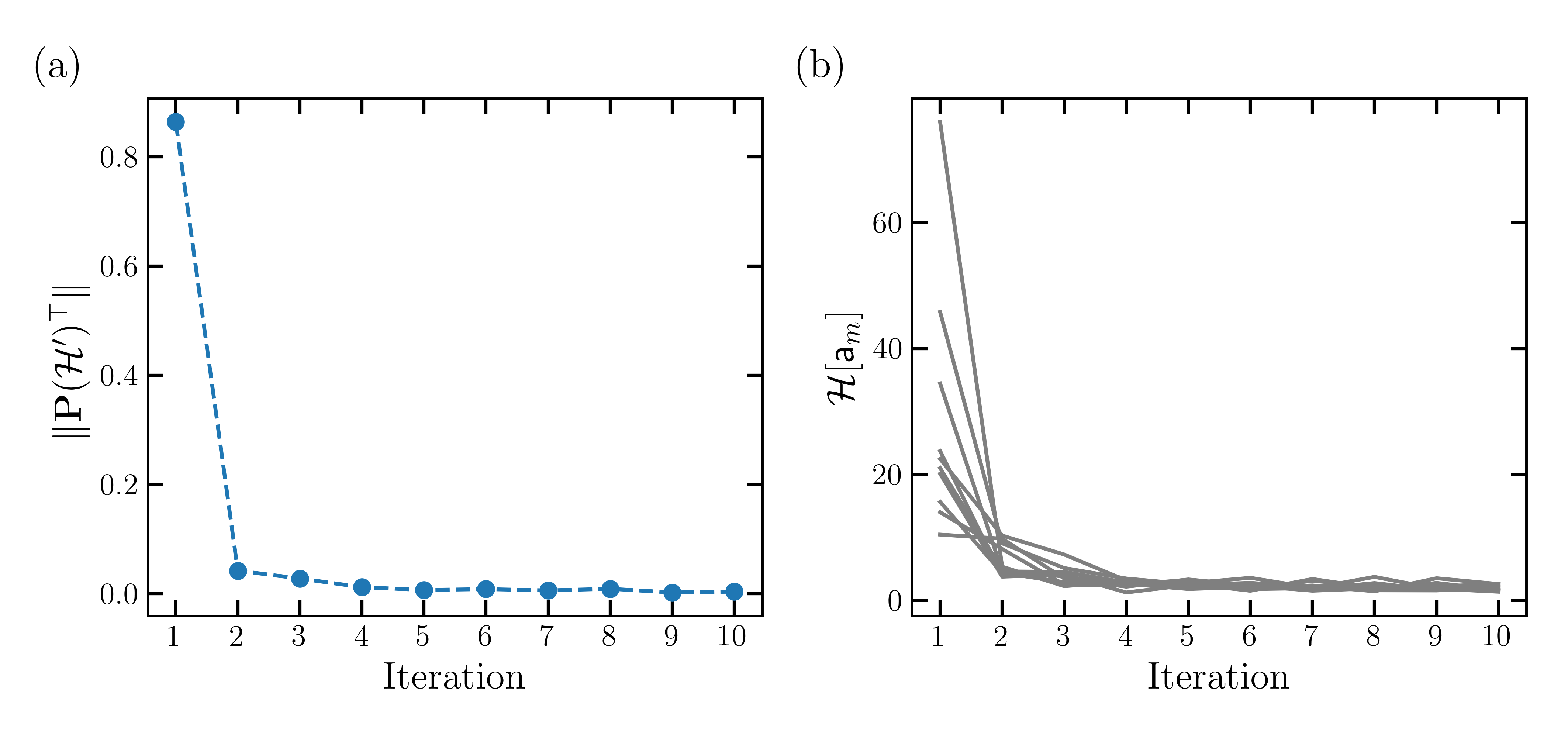}
    \caption{(a) The covariance-based gradient~$\mathbf{P}(\mathcal{H}')^\top$ and (b) the sample evaluations $\mathcal{H}[\mathsf{a}]$ at each iteration.}
    \label{fig:aHa}
\end{figure}

In the ensemble-based optimization, the random samples can influence the geometric update in two ways.
First, the samples can affect the gradient estimate that is used to guide the shape changes.
The ensemble Kalman method is a statistical inference approach, using the sample covariance~$\mathbf{P}(\mathcal{H}')^\top$ to estimate the gradient of the cost function at the sample mean.
The samples far away from the mean can noticeably increase the sample covariance in their direction.
Second, the objective evaluation~$\mathcal{H}[\mathsf{a}_m]$ of each sample determines the step length according to the Kalman-based update scheme.
For the sample~$\mathcal{H}[\mathsf{a}_m]$ away from zero, the step length is relatively large compared to other samples.
For the trailing edge optimization case in this work, the sample covariance and the sample evaluation at each iteration are presented in Figure~\ref{fig:aHa}.
It can be seen that the sample covariance is large at the initial step and reduces significantly at the second step.
This indicates the gradient estimate is relatively large for all samples at the first step and gradually vanishes to achieve the optimization convergence.
On the other hand, the sample evaluation shows that there are three samples larger than 30 and deviate far from zero compared to the rest samples.
These samples are reduced significantly at the second step, driven by the mean gradient and the relatively large evaluation.

\section{The regional influence of the integral source} 
\label{sec:source region}
\begin{figure}[!htb]
    \centering
    \includegraphics[width=.9\linewidth]{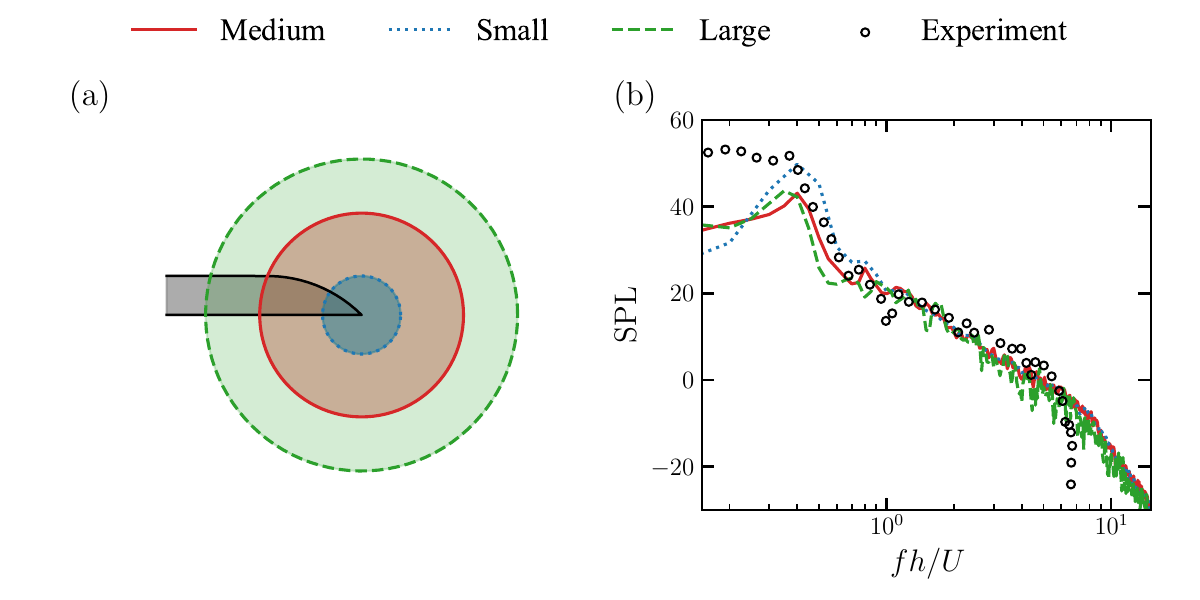}
    \caption{
    The influence of the domain size for the cylindrical integral source. (a) Schematic diagram of the cylindrical integral source region with different sizes. (b) Comparison of the far-field SPL.}

    \label{fig:intergal-size}
\end{figure}

Small regions of the integral source can better satisfy the theoretical assumptions of the Ffowcs Williams \& Hall acoustic analogy, given the hypothesis of compact noise sources.
On the other hand, the large region can ensure numerical convergence by encompassing sufficient decay of the Green's function, considering that the Green function decays as $r_0^{-3/2}$.
To choose an appropriate region size, we select three cylindrical integral regions with different radii~($R_0/h=1.0,2.6,4.0$) to compute the far-field sound pressure levels. 
The results are presented in Figure~\ref{fig:intergal-size}.
It can be seen that the SPLs with the three regions are consistent in the high-frequency range.
However, the SPL with the small region exhibits noticeable discrepancies at very low frequencies, while the large and medium regions yield similar results.  
Consequently, we choose the medium integral region to have a converged integral solution for the trailing edge noise in this work.

\section*{Acknowledgment}
This work is supported by NSFC Excellence Research Group Program for ``Multiscale Problems in Nonlinear Mechanics'' (No. 12588201). 
XLZ also acknowledges support from the Young Elite Scientists Sponsorship Program by CAST (No. 2022QNRC001) and the CAS Project for Young Scientists in Basic Research (YSBR-087).

\end{document}